\documentclass[fleqn,usenatbib]{mnras}

\usepackage{newtxtext,newtxmath}
\usepackage[T1]{fontenc}

\DeclareRobustCommand{\VAN}[3]{#2}
\let\VANthebibliography\thebibliography
\def\thebibliography{\DeclareRobustCommand{\VAN}[3]{##3}\VANthebibliography}

\usepackage{graphicx}	
\usepackage{amsmath}	
\usepackage{makecell}
\usepackage{multirow}
\usepackage{array}
\usepackage{tikz}
\usetikzlibrary{tikzmark}
\usepackage{orcidlink}

\newcommand{\numth}{$^{\rm th}~$}	%

\title[Direct chemical abundances in DESI DR2]{Electron temperature relations and the direct N, O, Ne, S and Ar abundances of 49$\,$959 star-forming galaxies in DESI Data Release 2}

\author[D. Scholte et al.]{
D.~Scholte$^{\orcidlink{0000-0002-6867-1244}}$,$^{1}$\thanks{E-mail: dscholte@ed.ac.uk (DS)}
F.~Cullen$^{\orcidlink{0000-0002-3736-476X}}$,$^{1}$
J.~Moustakas$^{\orcidlink{0000-0002-2733-4559}}$,$^{2}$
H.~Zou$^{\orcidlink{0000-0002-6684-3997}}$,$^{3}$
A.~Saintonge$^{\orcidlink{0000-0003-4357-3450}}$,$^{4,5}$
\newauthor{
K.~Z.~Arellano-Cordova$^{\orcidlink{0000-0002-2644-3518}}$,$^{1}$
T.~M.~Stanton$^{\orcidlink{0000-0002-0827-9769}}$,$^{1}$
B.~Andrews,$^{6}$ 
J.~Sui,$^{3}$ 
J.~Aguilar,$^{7}$ 
S.~Ahlen$^{\orcidlink{0000-0001-6098-7247}}$,$^{8}$
}
\newauthor{
D.~Bianchi$^{\orcidlink{0000-0001-9712-0006}}$,$^{9,10}$
D.~Brooks,$^{4}$ 
F.~J.~Castander$^{\orcidlink{0000-0001-7316-4573}}$,$^{11,12}$
T.~Cheng$^{\orcidlink{0000-0001-8670-4495}}$,$^{13}$
T.~Claybaugh,$^{7}$ 
}
\newauthor{
A.~de la Macorra$^{\orcidlink{0000-0002-1769-1640}}$,$^{14}$
B.~Dey$^{\orcidlink{0000-0002-5665-7912}}$,$^{15,6}$
P.~Doel,$^{4}$ 
K.~Douglass$^{\orcidlink{0000-0002-9540-546X}}$,$^{16}$
S.~Ferraro$^{\orcidlink{0000-0003-4992-7854}}$,$^{7,17}$
}
\newauthor{
J.~E.~Forero-Romero$^{\orcidlink{0000-0002-2890-3725}}$,$^{18,19}$
E.~Gaztañaga$^{\orcidlink{0000-0001-9632-0815}}$,$^{11,20,12}$
S.~Gontcho A Gontcho$^{\orcidlink{0000-0003-3142-233X}}$,$^{7,21}$
G.~Gutierrez,$^{22}$ 
}
\newauthor{
R.~Joyce$^{\orcidlink{0000-0003-0201-5241}}$,$^{23, \dagger}$
A.~Kremin$^{\orcidlink{0000-0001-6356-7424}}$,$^{7}$
O.~Lahav,$^{4}$ 
M.~Landriau$^{\orcidlink{0000-0003-1838-8528}}$,$^{7}$
L.~Le~Guillou$^{\orcidlink{0000-0001-7178-8868}}$,$^{24}$
P.~Martini$^{\orcidlink{0000-0002-4279-4182}}$,$^{25,26,27}$
}
\newauthor{
A.~Meisner$^{\orcidlink{0000-0002-1125-7384}}$,$^{23}$
R.~Miquel,$^{28,29}$ 
W.~J.~Percival$^{\orcidlink{0000-0002-0644-5727}}$,$^{30,31,32}$
C.~Poppett,$^{7,33,17}$ 
F.~Prada$^{\orcidlink{0000-0001-7145-8674}}$,$^{34}$
}
\newauthor{
I.~P\'erez-R\`afols$^{\orcidlink{0000-0001-6979-0125}}$,$^{35}$
G.~Rossi,$^{36}$ 
E.~Sanchez$^{\orcidlink{0000-0002-9646-8198}}$,$^{37}$
D.~Schlegel,$^{7}$ 
Z.~Shao,$^{38}$ 
J.~Silber$^{\orcidlink{0000-0002-3461-0320}}$,$^{7}$
}
\newauthor{
D.~Sprayberry,$^{23}$ 
G.~Tarl\'{e}$^{\orcidlink{0000-0003-1704-0781}}$$^{39}$
and B.~A.~Weaver$^{23}$ 
}
\\
\\
Author affiliations are listed at the end of the paper. $^{\dagger}$Deceased.
}

\date{Accepted XXX. Received YYY; in original form ZZZ}

\pubyear{\the\year{}}

\begin{document}
\label{firstpage}
\pagerange{\pageref{firstpage}--\pageref{lastpage}}
\maketitle

\begin{abstract}
We present the largest direct-method abundance catalogue of galaxies to date, containing measurements of 49$\,$959 star-forming galaxies at $z < 0.96$ from DESI Data Release 2. By directly measuring electron temperatures across multiple ionisation zones, we provide constraints on a number of electron temperature relations. Using the temperature measurements, we derive reliable abundances for N, O, Ne, S and Ar and measure the evolution of abundances and abundance ratios of as a function of metallicity and other galaxy properties. Our measurements include direct oxygen abundances for 49$\,$507 galaxies, leading to the discovery of the two most metal-poor galaxies in the nearby Universe, with oxygen abundances of $\rm 12+\log(O/H) = 6.77_{-0.03}^{+0.03}~\rm dex$ (1.2\% $\rm Z_{\odot}$) and $\rm 12+\log(O/H) = 6.81_{-0.04}^{+0.04}~\rm dex$ (1.3\% $\rm Z_{\odot}$). We identify a rare outlier population of 24 galaxies with high N/O ratios at low metallicity, reminiscent of galaxy abundances observed in the early Universe. We find the Ne/O ratio is constant at low metallicity but increases gradually at $\rm 12+log(O/H) > 8.105\pm0.004$ dex. We show that the S/O and Ar/O abundance ratios are strongly correlated, consistent with the expected additional Type Ia enrichment channel for S and Ar. In this work we present an initial survey of the key properties of the sample, with this dataset serving as a foundation for extensive future work on galaxy abundances at low redshift.
\end{abstract}




\section{Introduction}

The chemical enrichment of galaxies through nucleosynthesis in successive generations of stars is a key process in galaxy evolution. The vast majority of elements, except for hydrogen, helium and some lithium \citep{alpher1948}, are synthesised through various processes in different types of stars \citep{burbidge1957}. The relative abundance ratios of different elements in galaxies are a product of their respective production pathways in e.g., massive stars and intermediate mass stars or SNe Ia \citep[e.g.,][]{kobayashi2020} and therefore sensitive to the star-formation history. The gas-phase abundances of elements in galaxies are set through the interplay between their production, dilution through the inflows of pristine, unenriched gas and removal through processes such as stellar winds and supernovae; this complex interplay of processes is often referred to as the baryon cycle of galaxies \citep{tinsley1980}.

The gas phase abundances of N, O, Ne, S and Ar can be measured through their emission lines in the rest-frame optical spectra of star forming galaxies. The measurement of gas phase abundances has been an ongoing effort over successive generations of spectroscopic surveys \citep[see e.g.,][]{tremonti2004, izotov2006, guseva2011, curti2017, curti2020, berg2020, rogers2022, arellano-cordova2024, scholte2024, esteban2025, sanders2025}. Most studies focus on the abundance of oxygen, which is the most abundant metal in the ISM and therefore serves as a key tracer for understanding the chemical evolution of galaxies \citep[see][]{tremonti2004}. However, detailed studies of multiple abundances provide a more complete picture, since abundance ratios are sensitive to the different production pathways of elements. Elements with similar production pathways exhibit fairly constant abundance ratios over a wide range of metallicities (and other galaxy properties). This is the case for elements produced in the triple-alpha process in massive stars \citep[$M_{\star}>8\rm ~M_{\odot}$, see e.g.,][]{woosley2002}, so called alpha-elements such as O, Ne, S and Ar, of which the relative abundances are approximately constant \citep[e.g.][]{izotov2006, berg2019,arellano-cordova2024,esteban2025}. However, S and Ar each have a secondary production pathway through SNe Ia, due to which there is a delayed release of a significant fraction of the abundance of these elements \citep{matteucci2005, arnaboldi2022, rogers2024, stanton2025, bhattacharya2025, foley2025}. Elements with different production pathways, such as nitrogen, exhibit a more complex behaviour compared to alpha-elements. Nitrogen is primarily produced in intermediate mass stars (${\rm 4~M_{\odot}}<M_{\star}<8{\rm ~M_{\odot}}$), which release elements into the ISM on a much longer timescale with a typical lag time of approximately 250 Myr \citep[see][]{henry2000}. The N/O ratio is expected to be roughly constant at low metallicities, but to increase at higher metallicities \citep{izotov2006,nicholls2017, berg2012, berg2019,arellano-cordova2025}. The low-metallicity plateau is a result of the relatively low nitrogen production in massive stars, and the increasing trend with metallicity is the result of the delayed release of nitrogen from intermediate mass stars \cite[e.g.,][]{vanzee1998,nicholls2017}. However, detailed modelling of the chemical evolution of galaxies is required to reproduce the observed O/H versus N/O relation \citep[see][for detailed discussion on this]{vincenzo2018a, vincenzo2018b, kobayashi2020}. There has also been particular attention on a population of high N/O outliers mostly observed at high-redshift \citep[e.g.,][]{cameron2023,marques-chaves2024, topping2024}. The physical processes driving these unusual abundance patterns are still debated, however, it may be possible that short lived populations of Wolf-Rayet stars or even more massive stars are required to produce such abundance patterns \citep{berg2011, berg2025, vink2023}.

One of the most reliable methods to derive abundances from the emission lines of galaxy spectra is using the direct method \citep{peimbert1967}, which relies on the detection of faint auroral emission lines. This is in contrast to the main alternative metallicity measurements based solely on the strong emission lines in galaxy spectra, referred to as strong line calibrations, which are significantly less reliable metallicity diagnostics with different estimates varying up to 1 dex for individual galaxies \citep[][]{kewley2008, kewley2019}. Therefore, the direct method has become the "gold standard" for abundance measurements in H\textsc{ii}-regions and galaxies and is at the core of our understanding of the chemical evolution of galaxies \citep[see e.g.,][]{andrews2013, curti2020}. The reliability of the method is due to the ability to constrain abundances based only on atomic physics which determines the line emission from measurable gas properties such as electron density, $n_{\rm e}$, electron temperature, $T_{\rm e}$, and ionic abundances of various elements, X/H \citep[][]{aller1984}.

In most applications of the direct method, electron temperature relations between the electron temperatures of different ionic species account for the temperature structure of H\textsc{ii}-regions in abundance analyses \citep[e.g.,][]{dopita2000}. Detailed studies of the temperature structure show that ions are separated in several ionisation zones depending on their ionisation potential \citep[e.g.,][]{berg2021}. The electron temperature of each ion is either constrained directly or estimated from the temperature of a measured ionic state using electron temperature relations \citep[for discussions on electron temperature relations, see e.g.,][]{campbell1986,garnett1992,izotov2006,pilyugin2009,andrews2013, nicholls2014, croxall2016, arellano-cordova2020, rogers2021, rogers2022, mendez-delgado2023, cataldi2025}. Here, there is a difference between the electron temperature relations of individual H\textsc{ii}-regions and integrated galaxy spectra, where averaging over many H\textsc{ii}-regions with varying temperatures increases the scatter in galaxy integrated electron temperature relations. Additionally, due to the integration over the light of an entire galaxy there can be contamination from diffuse ionised gas \citep[DIG; see e.g.,][]{zhang2017, sanders2017, valeasari2019, mannucci2021, belfiore2022}. Several parametrisations accounting for or reducing the scatter in electron temperature relations have been proposed \citep[e.g.,][]{perez-montero2003, pilyugin2007, yates2020}. One of the key limitations on the derivation of electron temperature relations is the relatively small number of individual H\textsc{ii}-regions and galaxies with direct measurements of multiple auroral lines. 

In this study, we measure the fluxes of key emission lines to constrain the electron temperatures, densities and abundances of multiple elements using the direct method based on observations from the Dark Energy Spectroscopic Instrument \citep{levi2013,desi2023}. In particular, the increased depth in comparison to, e.g., the SDSS main galaxy sample \citep{york2000} and the large number of galaxies targeted over the wide survey footprint make it possible to detect auroral lines in an unprecedented number of galaxies \citep[see also][]{zou2024}. Pre-DESI studies typically contain hundreds of auroral line measurements of individual galaxies \citep[see e.g.,][]{izotov2006, curti2017, curti2020}. Therefore, DESI provides an increase of nearly two orders of magnitude in the number of auroral line detections of individual galaxies. We use these measurements to provide new constraints on electron temperature relations, verify ionisation correction factors and derive direct-abundances (of N, O, Ne, S and Ar). This includes several thousand galaxies with multiple electron temperature constraints, thousands of extremely metal-poor galaxies and other outlier populations.

The DESI survey is a wide-field spectroscopic survey that aims to measure the spectra of 63 million galaxies, quasars and stars over a period of eight years \citep{desi2023,desi2025dr1}. DESI is a robotic, multiplexed spectroscopic instrument on the Mayall 4-meter telescope at Kitt Peak National Observatory \citep{desi2022}. The DESI instrument obtains simultaneous spectra of almost 5000 objects \citep{desi2016b, silber2023, miller2023, poppett2024} and is currently conducting a survey of about a third of the sky \citep{desi2016, schlafly2023}, with the main purpose of using galaxies as cosmological probes \citep[see e.g.,][]{desi2024dr1cosmo,desidr22025bao}. The DESI survey contains 4 distinct target classes with different target selection criteria: the Bright Galaxy Survey \citep[BGS;][]{hahn2023bgs}, the Luminous Red Galaxy survey \citep[LRG;][]{zhou2023lrg}, the Emission Line Galaxy survey \citep[ELG;][]{raichoor2023elg} and the Quasar survey \citep[QSO;][]{chaussidon2023}. In this work the most important is the BGS survey which observes a magnitude limited sample of galaxies between $0.0<z<0.6$. There are also a number of secondary target programmes that are observed alongside the main survey programmes. In this work, the data observed as part of the LOWZ programme are particularly relevant; this programme targets low mass galaxies at low redshifts \citep{darragh-ford2023lowz}. 

This paper is structured as follows: in Section \ref{sec:observations} we describe the observations and data products used in this work including the basic analysis of the DESI spectra and sample selection. In Section \ref{sec:fiducial_model} we outline our fiducial model used for our abundance analysis. In Section \ref{sec:results} we present and discuss the results of our analysis, including the derived electron temperatures and abundances. Finally, in Section \ref{sec:summary} we summarise our findings and outline potential future work. We assume the cosmological parameters from the \cite{planck2020cmb}, a \cite{chabrier2003} initial mass function and solar abundances as in \cite{asplund2021}.

\section{Observations and data products} \label{sec:observations}
The galaxies in our sample are part of DESI data release 2 (DR2) which includes observations conducted in the survey validation period and the main survey between May 2021 and April 2024 (DESI collaboration et al. in prep.). The DESI spectrographs cover a wavelength range between 3600 \AA{} and 9800 \AA{} with a resolving power, $R = \lambda/\Delta\lambda$, ranging from 2000 at the shortest wavelengths to 5500 at the longest wavelengths \citep{desi2023}. The spectra are processed using an extensive spectroscopic reduction \citep{guy2023}, classification and redshifting pipeline \citep[][Bailey et al. in prep.]{anand2024}. The spectral energy distributions (SED) and emission lines of all DESI galaxies are modelled using the \textsc{FastSpecFit}\footnote{For a full overview of the \textsc{FastSpecFit} datamodel see: \\ \hyperlink{https://fastspecfit.readthedocs.io/en/latest/fastspec.html}{https://fastspecfit.readthedocs.io/en/latest/fastspec.html}} pipeline \citep[][]{moustakas2023}. In this work we use version 1.0 of the \textsc{FastSpecFit} catalogue of DR2 galaxies (loa-v1.0 catalogue). The DESI spectra and value added catalogues used in this work will be made publicly available in DESI public data release 2. Some of these data have already been made public as part of the DESI early data release \citep[EDR;][]{desi2023edr} and data release 1 \citep[DR1;][]{desi2024dr1cosmo, desi2025dr1}. DESI DR1 includes survey validation observations and main survey measurements obtained between May 2021 and June 2022.

The photometric measurements used in this work are derived from the 9\numth data release of the DESI Legacy Imaging Surveys \citep[][]{dey2019}. The photometry consists of three optical bands ($g$, $r$, $z$) observed using the Mayall $z$-band Legacy Survey (MzLS), Dark Energy Camera Legacy Survey (DECaLS), and Beijing-Arizona Sky Survey \citep[BASS;][]{zou2017} and observations in four infrared bands ($W1-4$) taken from the Wide-field Infrared Survey Explorer \citep[WISE;][]{mainzer2014}. We also make use of the Siena Galaxy Atlas \citep[SGA-2020;][]{moustakas2023sga} which is a catalogue of angular diameter-selected galaxies based on the DESI Legacy Survey imaging.

We select galaxies from the DESI DR2 dataset based on several criteria. We only select the best available spectra of each galaxy with a successful redshift measurement using the following selection in the DESI redshift catalogue: \texttt{ZCAT\_PRIMARY==True}, \texttt{SPECTYPE==GALAXY}, \texttt{ZWARN==0} and \texttt{DELTACHI2>=40}. The redshifts of DESI galaxies are measured using the RedRock template-fitting pipeline to derive classifications and redshifts for each targeted source \citep[][Bailey et al. in prep.]{anand2024}. The detection of faint auroral emission lines is crucial for the analysis in this work. Therefore, we perform a signal-to-noise ($\rm S/N$) selection where we only select galaxies with $\rm S/N>5$ detection of the line flux of at least one of the following auroral lines or doublets in the \textsc{FastSpecFit}
 emission line catalogue of the full DESI DR2 data: [\textsc{Oiii}]$\lambda$4363, [\textsc{Nii}]$\lambda$5755, [\textsc{Siii}]$\lambda$6312 or [\textsc{Oii}]$\lambda\lambda$7320,7330. We also require galaxies with $\rm EW(H\beta)_{rest}>20$, to ensure the galaxies in the sample have a good detection of emission lines that are not significantly affected by uncertainties in the continuum subtraction. This selection criterion also ensures that the selected galaxy sample is not strongly affected by emission line contamination from diffuse ionised gas \citep[DIG; see][]{valeasari2019, mannucci2021}.  

\begin{figure*}
	\begin{center}
	\includegraphics[width=0.99\textwidth]{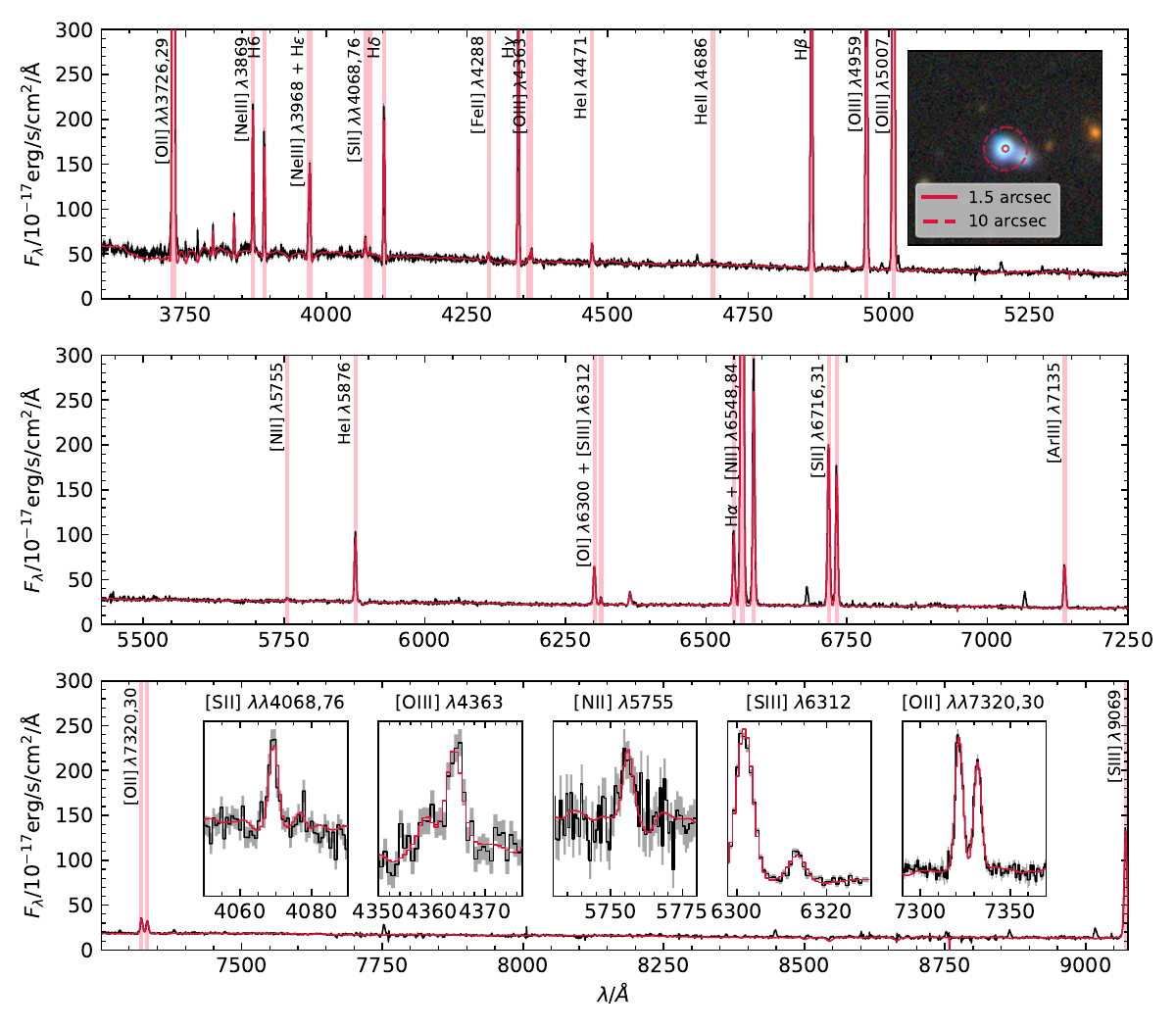}
	\caption{Figure of an example DESI spectrum of DESI J031.5272+08.5595 with the fitted \textsc{FastSpecFit} \citep{moustakas2023} model. The spectrum is shown in black with uncertainties in the flux measurements shown by the grey band. The fitted model is shown in red. We highlight the fitted emission lines in each panel with pink vertical bands and the names of the fitted lines. In the bottom panel we show five inset figures showing the fitted emission lines of the [\textsc{Sii}]$\lambda\lambda$4068,4076, [\textsc{Oiii}]$\lambda$4363, [\textsc{Nii}]$\lambda$5755, [\textsc{Siii}]$\lambda$6312 and [\textsc{Oii}]$\lambda\lambda$7320,7330 auroral lines. This spectrum was specifically chosen as all five auroral lines/doublets are detected. The inset image in the top panel shows the Legacy Survey imaging of this target with the 1.5" aperture of the DESI optical fiber at the pointing location of the observation as well as a 10" aperture for scale.}
	\label{fig:example_spectrum}
	\end{center}
\end{figure*}

The spectra of all the selected galaxies are refit with \textsc{FastSpecFit}, using a custom set of SED templates, a custom emission line list, and an improved uncertainty measurement using 100 Monte Carlo simulations of the fit to each spectrum. These updates tailor the SED fitting to the sample of galaxies selected here and ensure the reliable detection of particularly faint emission lines. The SED templates are derived using a \cite{chabrier2003} initial mass function, MIST isochrones \citep{choi2016} and FSPS population synthesis \citep{conroy2009, conroy2010}. We parametrise the star formation history using 8 variable-width age bins with constant star formation in each bin: 
$0-30$ Myr, $30-100$ Myr, $100-259$ Myr, $259-670$ Myr, $670$ Myr $-1.73$ Gyr, $1.73-4.50$ Gyr, $4.50-11.6$ Gyr and $11.6-13.7$ Gyr. We also include templates in three metallicity bins: 10\% $\rm Z_{\odot}$, 50\% $\rm Z_{\odot}$ and 100\% $\rm Z_{\odot}$. Dust emission is modelled using the \cite{draine2007} dust model where we have adopted the following values: $q_{\rm PAH}=1\%$, $U_{\rm min}=1$ and $\gamma=0.01$ \citep[see also][]{draine2007sings}. Our custom emission line list (see Table \ref{tab:linelist}) includes a large number of faint emission lines, e.g., we include measurements of the [\textsc{Sii}]$\lambda\lambda$4068, and the [Ar\textsc{IV}]$\lambda$4740 emission lines which are essential for our electron temperature and abundance measurements. The results from the custom \textsc{FastSpecFit} analysis are used for measurements of emission line fluxes and stellar masses used throughout this work. An example of a fit to a spectrum is shown in Figure \ref{fig:example_spectrum}.

\renewcommand{\arraystretch}{1.25}
\begin{table}
\centering
\caption{Custom emission line list.}
\begin{tabular}{lcl}
\hline
Line name & Rest wavelength & Comment \\
\hline \hline
[NeV] $\lambda$3346 & 3346.79 &  \\\relax
[NeV] $\lambda$3426 & 3426.85 &  \\\relax
[OII] $\lambda$3726 & 3727.10 &  \\\relax
[OII] $\lambda$3729 & 3729.86 &  \\\relax
H$\theta$ & 3798.978 &  \\\relax
H$\eta$ & 3836.478 &  \\\relax
[NeIII] $\lambda$3869 & 3869.86 &  \\\relax
H$\zeta$ & 3890.166 &  \\\relax
[NeIII] $\lambda$3968 & 3968.593 &  \\\relax
H$\epsilon$ & 3971.198 &  \\\relax
[SII] $\lambda$4068 & 4069.750 &  \\\relax
[SII] $\lambda$4076 & 4077.500 &  \\\relax
H$\delta$ & 4102.892 &  \\\relax
[FeII] $\lambda$4288 & 4288.599 &  \\\relax
H$\gamma$ & 4341.692 &  \\\relax
[FeII] $\lambda$4360 & 4359.387 &  \\\relax
[OIII] $\lambda$4363 & 4364.436 &  \\\relax
HeI $\lambda$4471 & 4472.735 &  \\\relax
HeII $\lambda$4686 & 4687.02 &  \\\relax
[ArIV] $\lambda$4713 & 4713.574 &  \\\relax
[ArIV] $\lambda$4740 & 4741.495 &  \\\relax
H$\beta$ & 4862.71 &  \\\relax
[OIII] $\lambda$4959 & 4960.295 & \makecell{Flux tied to [OIII] $\lambda$5007\\using the flux ratio from\\\cite{dimitrijevic2007}.} \\\relax
[OIII] $\lambda$5007 & 5008.240 &  \\\relax
[NII] $\lambda$5755 & 5756.191 &  \\\relax
HeI $\lambda$5876 & 5877.249 &  \\\relax
[OI] $\lambda$6300 & 6302.046 &  \\\relax
[SIII] $\lambda$6312 & 6313.81 &  \\\relax
[OI] $\lambda$6364 & 6365.535 &  \\\relax
[NII] $\lambda$6548 & 6549.861 & \makecell{Flux tied to [NII] $\lambda$6584\\using the flux ratio from\\\cite{dojcinovic2023}.} \\\relax
H$\alpha$ & 6564.60 &  \\\relax
[NII] $\lambda$6584 & 6585.273 &  \\\relax
[SII] $\lambda$6716 & 6718.294 &  \\\relax
[SII] $\lambda$6731 & 6732.674 &  \\\relax
[ArIII] $\lambda$7135 & 7137.77 &  \\\relax
[OII] $\lambda$7320 & 7321.94 &  \\\relax
[OII] $\lambda$7330 & 7332.21 &  \\\relax
[SIII] $\lambda$9069 & 9071.1 &  \\\relax
[SIII] $\lambda$9532 & 9533.2 & \\\hline
\end{tabular}
\label{tab:linelist}
\end{table}
\renewcommand{\arraystretch}{1.}

After refitting the spectra, we reapply the same quality cut on our final custom emission line catalog; we select galaxies with $\rm S/N>5$ detection of the line flux of at least one of the auroral lines or doublets, this time including the [\textsc{Sii}]$\lambda$4069 auroral line. We also exclude any auroral line measurements where $1.5 \times {A_{\rm line}}<|r^{90}_{\rm line}|$, where $r^{90}_{\rm line}$ is the 90$^{\rm th}$ percentile residuals within a 60 \AA{} window around the emission line and $A_{\rm line}$ is the amplitude of the emission line. This selection removes line fits in noisy spectra or spectra with sky subtraction or continuum subtraction issues. Additionally, we exclude auroral line measurements where the line centre falls within 10\AA{} of the edge of one of the DESI cameras. Faint emission lines can be affected by inaccurate measurements in these regions. We only include [\textsc{Oiii}]$\lambda$4363 measurements where the 1-$\sigma$ linewidth is less than 2\AA{} as [\textsc{Oiii}]$\lambda$4363 is difficult to deblend from [Fe\textsc{ii}]$\lambda$4360 when the emission lines are broader \citep{curti2017}. To ensure we can constrain dust attenuation, we require a S/N>3 detection of H$\beta$ and either H$\alpha$ or H$\gamma$. 

The bulk of the targets in our sample selection are star-forming galaxies, however, there are two other categories: galaxies with emission from active galactic nuclei (AGN) and individual \textsc{Hii}-regions of nearby galaxies. We identify AGN in the [\textsc{Nii}]-BPT \citep[][]{baldwin1981} emission line diagram as galaxies which are not identified as star-forming using the criterion defined by \cite{kewley2001} and also not star-forming based on the criterion by \cite{kauffmann2003}. At $z\gtrsim0.45$ the H$\alpha$ and [\textsc{Nii}]$\lambda$6584 emission lines are not observed by DESI, therefore at these redshifts we identify AGN using the mass-excitation diagram \citep[see equation 1 in][]{juneau2014}. We also exclude any galaxies which have broad H$\alpha$ or H$\beta$ lines with $\sigma^{\rm broad}_{\rm H\alpha}>1000 \rm ~km ~s^{-1}$ or $\sigma^{\rm broad}_{\rm H\beta}>1000 \rm ~km ~s^{-1}$. Additionally, we exclude galaxies where $\sigma^{\rm narrow}_{\rm H\beta} > 2\times\sigma^{\rm narrow}_{\rm [OIII]\lambda5007}$, this selection excludes galaxies where the broad and narrow components of the Balmer lines are not fitted correctly. This occurs almost exclusively at $z\gtrsim0.48$, where the H$\alpha$ emission line is not observed and therefore the narrow and broad components of Balmer lines are more difficult to constrain. 

We identify individual \textsc{Hii}-regions of nearby galaxies using the SGA-2020 \citep[][]{moustakas2023sga}. We mask targets as likely resolved \textsc{Hii}-regions when located within the radius of the major-axis of the 26 mag arcsec$^{-2}$ $r$-band isophote; we conservatively use the circular radius and not the full elliptical shape of the isophote. We do not mask targets where $z>0.15$ or where the distance of the targeted DESI fiber from the centre of the SGA-2020 galaxy is less than 1.5 arcsec. 

\begin{figure}
	\begin{center}
	\includegraphics[width=1.\columnwidth]{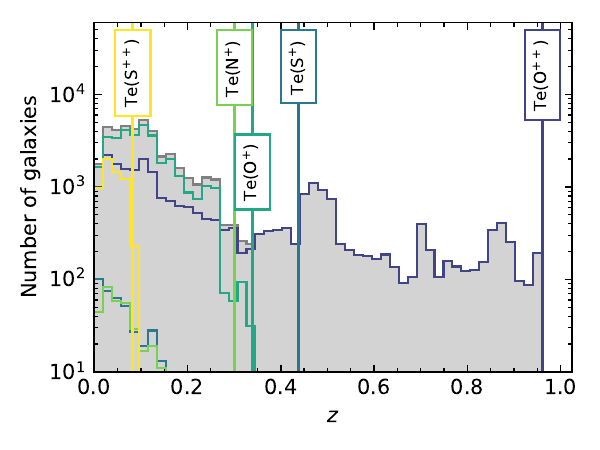}
	\caption{The redshift distribution of the galaxies with temperature measurements in our sample (grey). The vertical lines show the maximum redshift where a $T_{\rm e}$ has been determined based on the listed ionic species. The coloured histogram outlines in matching colours show the measurement distribution of electron temperatures of each ion.}
	\label{fig:redshift_dist}
	\end{center}
\end{figure}

Throughout this work we only include the data of star-forming galaxies, not including AGN or resolved \textsc{Hii}-regions of nearby galaxies. This final sample contains 49$\,$959 star-forming galaxies. The redshift distribution of our sample is shown in Figure \ref{fig:redshift_dist}.

\section{Fiducial model for nebular physical conditions and abundances} \label{sec:fiducial_model}
Our analysis of the nebular physical conditions and abundances is based on forward modelling using the \textsc{PyNeb} emission line analysis software \citep[][version 1.1.25]{pyneb2015}. Based on atomic physics, \textsc{PyNeb} allows us to derive model emission line flux ratios based on a wide range of ionised gas conditions in star-forming regions. The atomic data used in our analysis are summarised in Table \ref{tab:pyneb_data}. We use our forward model to infer the physical properties of the ionised gas that produces the emission lines observed in our spectra \citep[see also][]{scholte2025, cullen2025}. We sample the parameter-space using the \textsc{UltraNest} nested sampling algorithm \citep{buchner2021}. In this section, we describe the assumptions in our fiducial model, including electron temperature relations and ionisation correction factors (ICFs) used throughout the analysis. We have made choices that reflect some of the most common assumptions in the current literature. Therefore, our results can be readily compared to other literature studies. In Section \ref{sec:results}, we discuss these assumptions and compare to alternatives in more detail.

\renewcommand{\arraystretch}{1.25}
\begin{table}
\caption{Atomic data used in our analysis using \textsc{PyNeb}.}
\label{tab:pyneb_data}
\centering
\begin{tabular}{lll}
\hline
\textbf{Ion} & \textbf{Transition probabilities ($A_{\rm ij}$)} & \textbf{Collisional strengths ($Y_{\rm ij}$)} \\ \hline \hline
H$^{+}$ & \cite{storey1995} & --- \\
N$^{+}$ & \cite{froesefisher2004} & \cite{tayal2011} \\
O$^{+}$ & \cite{froesefisher2004} & \cite{kisielius2009} \\
O$^{++}$ & \cite{froesefisher2004} & \cite{storey2014} \\
Ne$^{++}$ & \cite{galavis1997} & \cite{bell2000} \\
S$^{+}$ & \cite{rynkun2019} & \cite{tayal2010} \\
S$^{++}$ & \cite{froesefisher2006} & \cite{tayal1999} \\
Ar$^{++}$ & \cite{munozburgos2009} & \cite{munozburgos2009} \\
Ar$^{3+}$ & \multirow{2}{10.5em}{\cite{mendoza1983},
    \cite{kaufman1986}} & \cite{ramsbottom1997} \\
    & & \\
\hline
\end{tabular}
\end{table}
\renewcommand{\arraystretch}{1.}

\renewcommand{\arraystretch}{1.25}
\begin{table*}
\caption{The observables, parameters and priors used in the \textsc{pyneb} analysis. The first step is to determine the electron density, temperature and dust attenuation, which are then used as priors in the second step to determine the ionic abundances. The priors are uniform distributions, $\mathcal{U}(a, b)$, where $a$ and $b$ are the lower and upper limits of the distribution.}
\label{tab:pyneb_table}
\centering
\begin{tabular}{c|c|cc}
\hline
    \multicolumn{1}{c}{\textbf{Observables}} & \multicolumn{1}{c}{\textbf{Parameters}} & \multicolumn{2}{c}{\textbf{Priors}} \\ \hline \hline
    \multicolumn{4}{c}{\textbf{Step 1:} electron density, temperature and dust attenuation} \\ \hline
    $\textsc{[Oii]}\lambda 3726/\textsc{[Oii]}\lambda 3729$ & $n_{\rm e}$\tikzmark{a} & $\mathcal{U}(10, 1000){~\rm cm^{-3}}$ & \\
    $\textsc{[Nii]}\lambda 5755/\textsc{[Nii]}\lambda 6584$ & $T_{\rm e}(\rm N^{+})$\tikzmark{i} & $\mathcal{U}(0.5, 3.5)\times10^4{~\rm K}$ & \\
    $\textsc{[Oii]}\lambda\lambda 7320,30/\textsc{[Oii]}\lambda\lambda 3726,29$ & $T_{\rm e}(\rm O^{+})$\tikzmark{g} & $\mathcal{U}(0.5, 3.5)\times10^4{~\rm K}$ &\\
    $\textsc{[Oiii]}\lambda 4363/\textsc{[Oiii]}\lambda 5007$ & $T_{\rm e}(\rm O^{++})$\tikzmark{c} & $\mathcal{U}(0.5, 3.5)\times10^4{~\rm K}$ & \\
    $\textsc{[Sii]}\lambda 4068/\textsc{[Sii]}\lambda\lambda 6716,31$ & $T_{\rm e}(\rm S^{+})$ & $\mathcal{U}(0.5, 3.5)\times10^4{~\rm K}$ & \\
    $\textsc{[Siii]}\lambda 6312/\textsc{[Siii]}\lambda 9069$ & $T_{\rm e}(\rm S^{++})$\tikzmark{e} & $\mathcal{U}(0.5, 3.5)\times10^4{~\rm K}$ & \\
    ${\rm H}\beta /{\rm H}\alpha$ & \multirow{2}{*}{$A_{\rm V}$} & \multirow{2}{*}{$\mathcal{U}(0.0, 4.0){~\rm mag}$} & \\
    ${\rm H}\gamma /{\rm H}\beta$ & & & \\ \hline
    \multicolumn{4}{c}{\textbf{Step 2:} ionic abundances} \\ \hline
    $\textsc{[Nii]}\lambda 6584/{\rm H}\beta$ & ${\rm log(N^{+}/H^{+})}$ & $\mathcal{U}(-9, -2)$ dex  & \multirow{8}{*}{{\Large+}
    $\begin{array}{c}
    {\rm Using ~posterior~distributions}\\
    {\rm determined~in ~step~1}\\
    \left(
\begin{array}{c}
    n_{\rm e}\\
    T_{\rm high}\\
    T_{\rm mid}\\
    T_{\rm low}\\
    A_{\rm V}
\end{array}
\right)\end{array}$
}\\
    $\textsc{[Oii]}\lambda\lambda 3726,3729/{\rm H}\beta$ & ${\rm log(O^{+}/H^{+})}$ & $\mathcal{U}(-9, -2)$ dex & \\
    $\textsc{[Oiii]}\lambda 5007/{\rm H}\beta$ & ${\rm log(O^{2+}/H^{+})}$ & $\mathcal{U}(-9, -2)$ dex & \\
    $\textsc{[}{\rm Ne}\textsc{iii]}\lambda 3869/{\rm H}\beta$ & ${\rm log(Ne^{2+}/H^{+})}$ & $\mathcal{U}(-9, -2)$ dex & \\
    $\textsc{[Sii]}\lambda\lambda 6716,6731/{\rm H}\beta$ & ${\rm log(S^{+}/H^{+})}$ & $\mathcal{U}(-9, -2)$ dex & \\
    $\textsc{[Siii]}\lambda 9069/{\rm H}\beta$ & ${\rm log(S^{2+}/H^{+})}$ & $\mathcal{U}(-9, -2)$ dex & \\
    $\textsc{[}{\rm Ar}\textsc{iii]}\lambda 7135/{\rm H}\beta$ & ${\rm log(Ar^{2+}/H^{+})}$ & $\mathcal{U}(-9, -2)$ dex & \\
    $\textsc{[}{\rm Ar}\textsc{iv]}\lambda 4740/{\rm H}\beta$ & ${\rm log(Ar^{3+}/H^{+})}$ & $\mathcal{U}(-9, -2)$ dex & \\ \hline
\end{tabular}
\end{table*}
\renewcommand{\arraystretch}{1.}

\subsection{Nebular physical conditions}
The forward modelling analysis is split into two steps. In the first step, we jointly derive the nebular physical conditions: electron density, $n_{\rm e}$, electron temperature, $T_{\rm e}$, and dust attenuation, $A_{\rm V}$. In the top half of Table \ref{tab:pyneb_table} we show the observable line ratios (column 1) used to constrain these parameters (column 2) together with the priors (column 3) placed on the physical parameters. We use non-informative, uniform priors for all parameters. The electron density is constrained using the [\textsc{Oii}]$\lambda$3726/[\textsc{Oii}]$\lambda$3729 doublet ratio which we choose over the [\textsc{Sii}]$\lambda$6731/[\textsc{Sii}]$\lambda$6716 ratio as the [\textsc{Sii}]-doublet is not in the observable wavelength range beyond $z\sim 0.45$, whereas the [\textsc{Oii}]-doublet is available for our full sample. As shown in the \textbf{top} half of Table \ref{tab:pyneb_table}, we have measurements of auroral lines from N$^{+}$, O$^{+}$, O$^{++}$, S$^{+}$ and S$^{++}$ ions, whenever detected with sufficient $\rm S/N$. We only report electron temperatures where there is a $\rm S/N>3$ detection\footnote{Our sample selection requires the detection of at least one of the auroral lines in a spectrum at $\rm S/N>5$, however, any subsequent auroral lines are still included in our analysis if $\rm S/N>3$. The more stringent sample selection criterion is to remove false positive auroral line detections. However, when one auroral line has reliably been detected at $\rm S/N>5$ the chance of spurious subsequent auroral lines is very low, and therefore a lower threshold is justified.} of the auroral line (e.g., [\textsc{Oiii}]4363) and a $\rm S/N>3$ detection of the accompanying strong emission line of the adjacent transition of the same species (e.g., [\textsc{Oiii}]5007 in case of the [\textsc{Oiii}]4363 auroral line). Dust attenuation is jointly constrained through the H$\beta$/H$\alpha$ and H$\gamma$/H$\beta$ line ratios using a \cite{cardelli1989} attenuation prescription with $R_{\rm V} = 3.1$ assuming case B recombination \citep[][]{baker1938}.

\subsection{Abundances}
Once the physical conditions are constrained, we derive ionic abundances of the $\rm N^{+}$, $\rm O^{+}$, $\rm O^{2+}$, $\rm Ne^{2+}$, $\rm S^{+}$, $\rm S^{2+}$ and $\rm Ar^{2+}$ and $\rm Ar^{3+}$ ions. We assume a three-zone temperature structure of the ionised gas. Each ion occupies a ionisation zone: $\rm O^{2+}$, $\rm Ne^{2+}$ and $\rm Ar^{3+}$ are in the high ionisation zone ($T_{\rm high}$); $\rm S^{2+}$ and $\rm Ar^{2+}$ are in the intermediate ionisation zone ($T_{\rm mid}$); $\rm N^{+}$, $\rm O^{+}$ and $\rm S^{+}$ are in the low ionisation zone ($T_{\rm low}$) \citep{mingozzi2022,berg2022}. The temperature of each zone can be constrained directly if the auroral line ratio of an ion in this zone is detected. Therefore, in this work $T_{\rm high}$ is constrained by $T_{\rm e}(\rm O^{++})$ and $T_{\rm mid}$ by $T_{\rm e}(\rm S^{++})$. In the low ionisation zone we have three possible constraints. We assume $T_{\rm low}$ is given by $T_{\rm e}(\rm O^{+})$ as this line ratio is detected for the majority of observations, otherwise we use $T_{\rm e}(\rm S^{+})$ or $T_{\rm e}(\rm N^{+})$. If the temperature in an ionisation zone is not constrained, then we infer the temperature from a different zone using electron temperature relations ($T_{\rm e}-T_{\rm e}$). We assume the temperature relation defined by \cite{campbell1986} and \cite{garnett1992} based on H\textsc{ii}-region models by \cite{stasinska1982} which provides a relation between the low- and high-ionisation zones: $T_{\rm low} = 0.7\times T_{\rm high}+3000 \rm K$. We assume the relation defined by \cite{croxall2016} for the relation between the intermediate- and high-ionisation zones: $T_{\rm mid} = 1.265\times T_{\rm high}-2320 \rm K$. A more detailed analysis of $T_{\rm e}-T_{\rm e}$ relations is provided in Section \ref{sec:tt_relations}. We constrain the ionic abundances of elements using the ratios of the respective strong emission line strength compared to H$\beta$ combined with the posterior distributions (derived in step 1) on the electron density, electron temperature in each zone and dust attenuation. The line ratios used to constrain the abundance of each ion are listed in Table \ref{tab:pyneb_table}. We only report ionic abundances where there is a reliable temperature measurement (based on the S/N criteria of the relevant emission lines) and the strong emission lines of the ion are detected at $\rm S/N>3$. 

The total oxygen abundances are derived using the sum of the abundances of the singly ($\rm O^{+}/H^{+}$) and doubly ($\rm O^{++}/H^{+}$) ionised states. The contribution of higher ionised states is considered to be negligible, even at high electron temperatures the contribution of this state is small \citep[e.g.,][]{berg2021, rickardsvaught2025, cullen2025}. To calculate the total abundances of each other element we use the ionisation correction factors by \cite{izotov2006}. These ionisation correction factors allow us to calculate the total abundances from measurements of an incomplete subset of ionic abundances. For N and Ne we only measure one ionic abundance each ($\rm N^{+}/H^{+}$ and $\rm Ne^{++}/H^{+}$, respectively). Therefore, the total abundances are estimated only from these single ionic abundances \citep[equations 18 and 19 in][respectively]{izotov2006}. For S we measure both the singly ($\rm S^{+}/H^{+}$) and doubly ($\rm S^{++}/H^{+}$) ionised state; we only calculate the total S abundances when both of these ionic abundances are constrained \citep[equation 20 in][]{izotov2006}. For Ar we measure the doubly ($\rm Ar^{++}/H^{+}$) and triply ($\rm Ar^{3+}/H^{+}$) ionised states. We calculate the total Ar abundance using both ionised states where each is constrained \citep[equation 23 in][]{izotov2006}; however, if $\rm Ar^{3+}/H^{+}$ is not constrained, we calculate the total Ar abundance from just the $\rm Ar^{++}/H^{+}$ ionic abundances \citep[equation 22 in][]{izotov2006}. We discuss the validity of the chosen ICFs in more detail in Section \ref{sec:icfs}. The total number of galaxies with abundance measurement of each ion/element is listed in Table \ref{tab:sample_numbers}. Nitrogen, oxygen, Neon and Argon abundances are each available for the majority of galaxies with an electron temperature measurement. The sulfur abundances are only available for a smaller subset as the [\textsc{Siii}]$\lambda$9069 emission line is only observable in DESI spectra at $z \lesssim0.08$.

\subsection{Star formation rates}
We derive the star formation rates from the aperture corrected and dust corrected H$\alpha$ flux (or H$\beta$ flux if H$\alpha$ is not available). We use the aperture corrections derived using \textsc{FastSpecFit} in Section \ref{sec:observations}. Our star-formation rates are calculated using the metallicity-dependent calibration by \cite{korhonencuestas2025} which is defined as
\begin{equation}
	\log\left(\frac{\dot{M}_{\star}}{{\rm M}_{\odot}~{\rm yr}^{-1}}\right) = \log\left(\frac{\mathcal{L}_{\rm H\alpha}}{\textrm{erg~s}^{-1}}\right) - \log(C(Z_{\star}/{\rm Z}_{\odot})),
\end{equation}
where $\mathcal{L}_{\rm H\alpha}$ is the dust corrected H$\alpha$ luminosity and $C(Z_{\star}/{\rm Z}_{\odot})$ is the metallicity-dependent conversion factor. The conversion factor is defined as outlined in section 2.4 of \cite{korhonencuestas2025}. We infer the stellar metallicity, $Z_{\star}/{\rm Z_{\odot}}$, from the gas phase oxygen abundances where we assume $\log(Z_{\star}/{\rm Z_{\odot}})=\rm \log(O/H)-\log(O/H)_{\odot}$. This assumption implies that there is no enhancement of alpha-elements as the stellar metallicity is defined in terms of iron abundances. At low-metallicity, the luminosity to SFR conversion factor of this calibration is lower than a typical conversion factor not accounting for metallicity \citep{kennicutt1998,hao2011}.

\section{Results and discussion} \label{sec:results}
In this work we present a large sample of electron temperature measurements and abundance measurements of N, O, Ne, S, Ar in star-forming galaxies. In this section, we present our results, and discuss implications and connections to previous work. The DESI DR2 dataset provides an unprecedented number of auroral line detections in star-forming galaxies. Due to this large sample size we are not only able to derive the average relations between electron temperatures and abundances, but also to explore the scatter and trends in more detail. The large sample size means we are more likely to detect outlier populations with rare properties. In this work we focus on analysing the electron temperature relations and provide a cursory overview of our abundance measurements. A more detailed analysis of the abundance measurements and their implications for the chemical evolution of galaxies will be presented in a follow-up paper (Scholte et al. in prep.). The total number of galaxies for which each measurement is available is summarised in Table \ref{tab:sample_numbers}.

\subsection{Electron temperature measurements and temperature relations}
\label{sec:tt_relations}
\begin{figure}
	\begin{center}
	\includegraphics[width=1.\columnwidth]{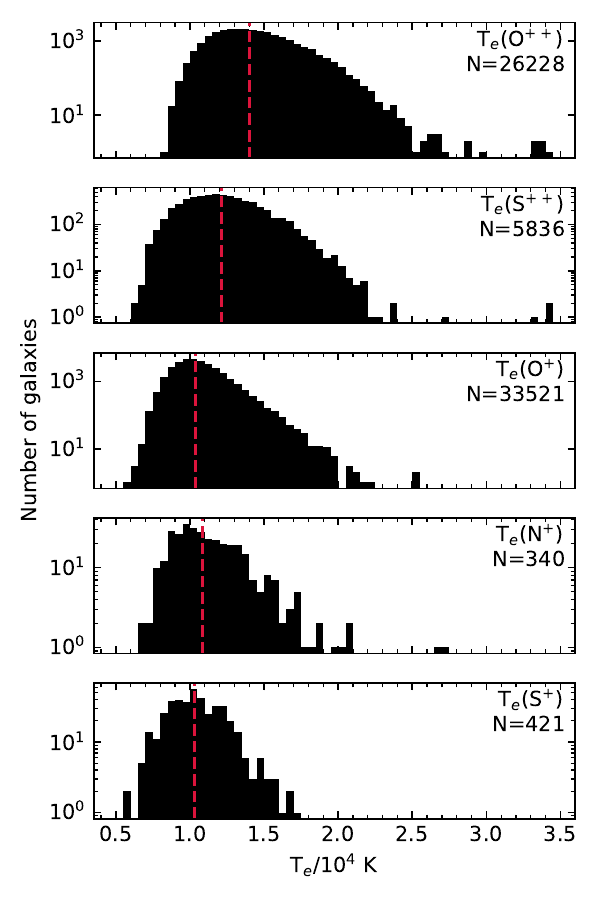}
	\caption{The histograms of the electron temperature measurements derived from the auroral emission lines of different ions. The red dashed lines show the median electron temperature measured using each auroral line. The [\textsc{Oiii}]$\lambda$4363 auroral line (corresponding to $T_{\rm e}\rm (O^{++})$) is brightest at high temperatures, the [\textsc{Siii}]$\lambda$6312 emission line (correcponding to $T_{\rm e}\rm (S^{++})$) is brightest at intermediate temperatures and $T_{\rm e}\rm (O^{+})$, $T_{\rm e}\rm (N^{+})$ and $T_{\rm e}\rm (S^{+})$ are brightest at low temperatures. The number of objects with temperature constraints using each ion are indicated in the top-right corner of each panel.}
	\label{fig:te_histograms}
	\end{center}
\end{figure}

We derive electron temperature measurements using the emission lines of 5 different ions with varying ionisation potentials. Figure \ref{fig:te_histograms} shows the histograms of the electron temperature measurements derived from each of these ions. The temperature range where we can detect these auroral emission lines is different for each species. The brightness of the line is a combination of the line emissivity (temperature and density dependent) and the abundance of the element. The combination of these factors results in the [\textsc{Oiii}]$\lambda$4363 auroral line (corresponding to $T_{\rm e}\rm (O^{++})$) being brightest at relatively high temperatures. The [\textsc{Siii}]$\lambda$6312 emission line (corresponding to $T_{\rm e}\rm (S^{++})$) is brightest at intermediate temperatures and [\textsc{Nii}]$\lambda$5755, [\textsc{Oii}]$\lambda\lambda$7320,7330 and [\textsc{Sii}]$\lambda\lambda$4068,4076 (corresponding to $T_{\rm e}\rm (N^{+})$, $T_{\rm e}\rm (O^{+})$ and $T_{\rm e}\rm (S^{+})$, respectively) are brightest at lower temperatures. Due to this fact, these auroral emission lines are detected in different subsets of galaxies. Additionally, due to temperature inhomogeneities in/between star-forming regions of galaxies, the temperatures measured using these different probes deviate when multiple auroral lines are detected in a galaxy. These effects are reflected in the median electron temperatures measured using each auroral line (red dashed lines in Fig. \ref{fig:te_histograms}). However, it is important to keep in mind that these electron temperatures are also measured over different redshift ranges. The histograms also show that, whilst our models allow electron temperatures up to 35,000 K, electron temperatures above $\sim$25,000 K are extremely rare in star-forming galaxies ($\sim0.05\%$ in our sample). 

We investigate the relations between the electron temperature measurements of different probes when multiple temperature measurements are available for any given galaxy. We fit temperature relations using orthogonal distance regression as implemented in \textsc{Scipy}  \citep[ODR;][]{boggs1981,scipy2020}. We include all temperature measurements of star-forming galaxies where the respective auroral and strong emission lines are detected at $\rm S/N>3$ and where $T_{\rm e}^{84}-T_{\rm e}^{16}<5000 \rm ~K$ (where $T_{\rm e}^{x}$ is the x\numth percentile of the posterior distribution of $T_{\rm e}$). The uncertainty in the $T_{\rm e}$ measurements are included as the weights, $w$, in the ODR fitting, where $w=((T_{\rm e}^{84}-T_{\rm e}^{16})/2)^{-2}$. We measure the scatter around the fitted relations as $\sigma_{\rm tot} = (r^{84}-r^{16})/2$ for residuals, $r$. The intrinsic scatter is measured as $\sigma_{\rm int} = \sqrt{\sigma_{\rm tot}^{2} - \widetilde{\sigma}_{\rm meas}^2}$, where $\widetilde{\sigma}_{\rm meas}$ is the median propagated measurement uncertainty in the residuals.

\renewcommand{\arraystretch}{1.25}
\begin{table}
\centering
\caption{Summary of electron temperature and abundance measurements from the \textsc{PyNeb} analysis. The table lists the parameters and the number of galaxies for which each parameter was measured. $^{\dagger}$The total number of galaxies for which at least one electron temperature was measured.}
\begin{tabular}{lr}
\hline
\multicolumn{2}{c}{\textbf{Electron temperature measurements}}\\ 
Parameter & Number of galaxies \\ \hline \hline
$T_{\rm e}\rm (N^{+})$ & 340 \\
$T_{\rm e}\rm (O^{+})$ & 33$\,$521 \\
$T_{\rm e}\rm (O^{++})$ & 26$\,$228 \\
$T_{\rm e}\rm (S^{+})$ & 421 \\
$T_{\rm e}\rm (S^{++})$ & 5$\,$836 \\ 
$T_{\rm e}$ & 49$\,$959$^{\dagger}$ \\ \hline
\multicolumn{2}{c}{\textbf{Abundance measurements}}\\ 
Parameter & Number of galaxies \\ \hline \hline
$\rm log(N^{+}/H^{+})$ & 42$\,$173 \\
$\rm log(N/H)$ & 42$\,$151 \\
$\rm log(O^{+}/H^{+})$ & 49$\,$507 \\
$\rm log(O^{++}/H^{+})$ & 49$\,$698 \\
$\rm log(O/H)$ & 49$\,$507 \\
$\rm log(Ne^{++}/H^{+})$ & 47$\,$719 \\
$\rm log(Ne/H)$ & 47$\,$529 \\
$\rm log(S^{+}/H^{+})$ & 41$\,$141 \\
$\rm log(S^{++}/H^{+})$ & 15$\,$778 \\
$\rm log(S/H)$ & 15$\,$585\\
$\rm log(Ar^{++}/H^{+})$ & 39$\,$739\\
$\rm log(Ar^{3+}/H^{+})$ & 1$\,$260 \\
$\rm log(Ar/H)$ & 39$\,$692 \\ \hline
\end{tabular}
\label{tab:sample_numbers}
\end{table}
\renewcommand{\arraystretch}{1.}

\begin{figure}
	\begin{center}
	\includegraphics[width=\columnwidth]{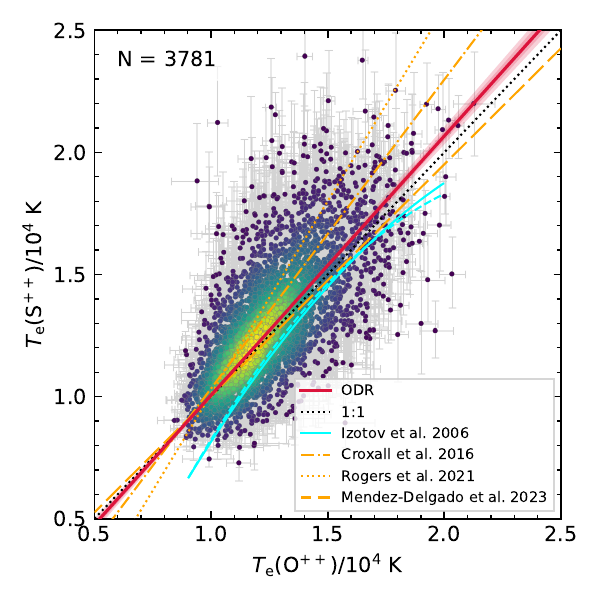}
	\caption{The electron temperature relation between the high-ionisation zone, $T_{\rm e}(\mathrm{O}^{++})$, and the intermediate-ionisation zone, $T_{\rm e}(\mathrm{S}^{++})$. The red line shows the best fit relation derived from our observations. The shaded region shows the 1$\sigma$ uncertainty in the relation. The total scatter in the relation is $\sigma_{\rm tot}(\mathrm{S}^{++}) =1700~\mathrm{K}$ and the intrinsic scatter is $\sigma_{\rm int}(\mathrm{S}^{++})=900~\mathrm{K}$. The total scatter in $T_{\rm e}(\mathrm{O}^{++})$ is $\sigma_{\rm tot}(\mathrm{O}^{++}) = 1600~\mathrm{K}$ and with intrinsic scatter $\sigma_{\rm int}(\mathrm{O}^{++})=900~\mathrm{K}$. The literature relations by \protect\cite{izotov2006}, \protect\cite{croxall2016} and \protect\cite{rogers2021} and \protect\cite{mendez-delgado2023} are shown for comparison as indicated in the legend. All data points are coloured by the Gaussian kernel density of the plotted distribution.}
	\label{fig:t_oiii_siii}
	\end{center}
\end{figure}

\begin{figure}
	\begin{center}
	\includegraphics[width=\columnwidth]{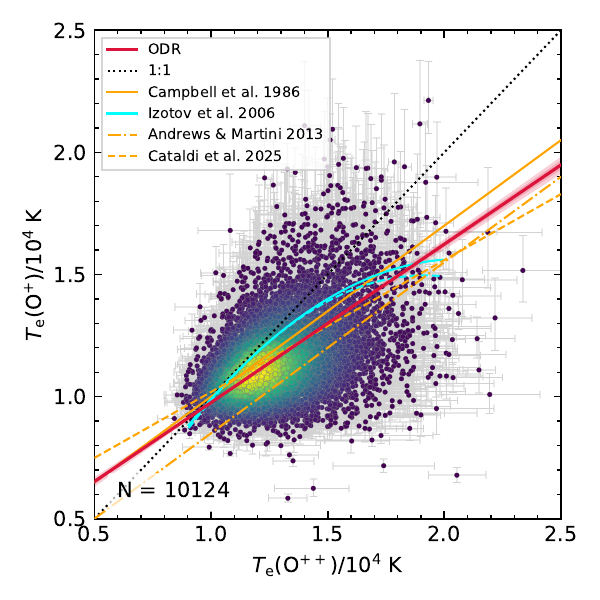}
	\caption{The electron temperature relation between the high-ionisation zone, $T_{\rm e}(\mathrm{O}^{++})$, and the low-ionisation zone, $T_{\rm e}(\mathrm{O}^{+})$. The red line shows the best fit relation derived from our observations. The shaded region shows the 1$\sigma$ uncertainty in the relation. The total scatter in the relation is $\sigma_{\rm tot}(\mathrm{O}^{+}) =2100~\mathrm{K}$ and the intrinsic scatter is $\sigma_{\rm int}(\mathrm{O}^{+})=1700~\mathrm{K}$. The total scatter in $T_{\rm e}(\mathrm{O}^{++})$ is smaller at $\sigma_{\rm tot}(\mathrm{O}^{++}) = 1400~\mathrm{K}$ and with intrinsic scatter $\sigma_{\rm int}(\mathrm{O}^{++})=1100~\mathrm{K}$. The literature relations by \protect\cite{campbell1986}, \protect\cite{izotov2006}, \protect\cite{andrews2013} and \protect\cite{cataldi2025} are shown for comparison as indicated in the legend. All data points are coloured by the Gaussian kernel density of the plotted distribution.}
	\label{fig:t_oiii_oii}
	\end{center}
\end{figure}

\begin{figure}
	\begin{center}
	\includegraphics[width=\columnwidth]{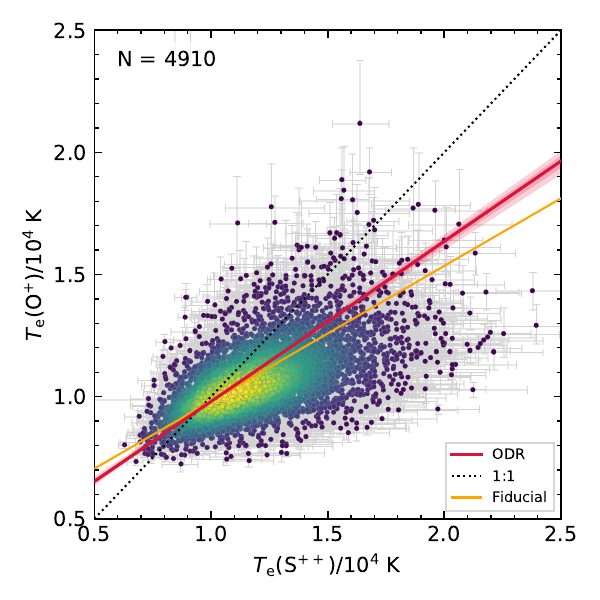}
    \caption{The electron temperature relation between the intermediate-ionisation zone, $T_{\rm e}(\mathrm{S}^{++})$, and the low-ionisation zone, $T_{\rm e}(\mathrm{O}^{+})$. The red line shows the best fit relation derived from our observations. The orange line is our fiducial model based on the \protect\cite{campbell1986} and \protect\cite{croxall2016} relations. The shaded region shows the 1$\sigma$ uncertainty in the relation. The total scatter in the relation is $\sigma_{\rm tot}(\mathrm{O}^{+}) = 1400~\mathrm{K}$ and the intrinsic scatter is $\sigma_{\rm int}(\mathrm{O}^{+})=1000~\mathrm{K}$. The total scatter in $T_{\rm e}(\mathrm{S}^{++})$ is larger at $\sigma_{\rm tot}(\mathrm{S}^{++}) = 2100~\mathrm{K}$ and with intrinsic scatter $\sigma_{\rm int}(\mathrm{S}^{++})=1600~\mathrm{K}$. All data points are coloured by the Gaussian kernel density of the plotted distribution.}
	\label{fig:t_siii_oii}
	\end{center}
\end{figure}

\subsubsection{The $T_{\rm e}(\rm O^{++})$ versus $T_{\rm e}(\rm S^{++})$ relation}
The constraints on the temperatures in the high- and intermediate-ionisation zones are constrained by the $T_{\rm e}(\rm O^{++})$ and $T_{\rm e}(\rm S^{++})$ measurements, respectively. In Figure \ref{fig:t_oiii_siii}, we show the measured electron temperatures and the relation derived from our observations, which is given by
\begin{equation}
    T_{\rm e}({\rm S^{++}}) = (1.062\pm0.022) \times T_{\rm e}({\rm O^{++}}) -(0.06\pm0.03) \times 10^4~\rm K.
\end{equation}
The total scatter in the relation is $\sigma_{\rm tot}(\rm S^{++}) =1700~\rm K$ and the intrinsic scatter is $\sigma_{\rm int}(\rm S^{++})=900~\rm K$, where the residual is defined as $T_{\rm e}(\rm S^{++})_{\rm obs} - T_{\rm e}(\rm S^{++})_{\rm inferred}$. The total scatter in $T_{\rm e}(\rm O^{++})$ is $\sigma_{\rm tot}(\rm O^{++}) = 1600~\rm K$ and with intrinsic scatter $\sigma_{\rm int}(\rm O^{++})=900~\rm K$. The average temperature relation we derive is very close to a one-to-one relation, showing that the $\rm O^{++}$ and $\rm S^{++}$ ions probe similar ionisation zones within the ionised gas in individual galaxies. Our electron temperature relation is less steeply increasing than the relations derived by \cite{croxall2016} and \cite{rogers2021}. Simultaneously, our $T_{\rm e}({\rm S^{++}})$ are slightly higher at fixed $T_{\rm e}({\rm {O++}})$ than expected from the relations by \cite{izotov2006}, based on photoionisation models, and \cite{mendez-delgado2023}, based on individual H\textsc{ii}-regions. The differences are smallest in comparison to the relation derived by \cite{croxall2016} which we used in our fiducial model and the relation derived by \cite{mendez-delgado2023}. Some differences we observe might also be due to the fact that the relations by \cite{croxall2016}, \cite{rogers2021} and \cite{mendez-delgado2023} are derived based on individual H\textsc{ii}-regions and our measurements are of integrated galaxy spectra. The intrinsic scatter in our temperature relations is significantly larger than the scatter in the relations derived through individual H\textsc{ii}-region measurements. This is most likely due to the integrated galaxy measurements, which leads to the blending of line emission from multiple H\textsc{ii}-regions with a range of physical properties.

\subsubsection{The $T_{\rm e}(\rm O^{++})$ versus $T_{\rm e}(\rm O^{+})$ relation}
The relation between $T_{\rm e}(\rm O^{++})$ and $T_{\rm e}(\rm O^{+})$ has been widely reported in previous literature \citep[e.g.,][]{campbell1986, garnett1992, izotov2006,andrews2013, yates2020, cataldi2025}. In Figure \ref{fig:t_oiii_oii} we show the measured electron temperatures and the relation derived from our observations, which is given by 
\begin{equation}
    T_{\rm e}({\rm O^{+}}) = (0.648\pm0.015) \times T_{\rm e}({\rm O^{++}}) + (0.327\pm0.018) \times 10^4~\rm K,
\end{equation}
with a total scatter of $\sigma_{\rm tot}(\rm O^{+}) =2100~\rm K$ and an intrinsic scatter of $\sigma_{\rm int}(\rm O^{+})=1700~\rm K$. The total scatter in $T_{\rm e}(\rm O^{++})$ is smaller at $\sigma_{\rm tot}(\rm O^{++}) = 1400~\rm K$ and with intrinsic scatter $\sigma_{\rm int}(\rm O^{++})=1100~\rm K$. Our relation differs slightly from the relation derived by \cite{campbell1986} and \cite{garnett1992} based on models by \cite{stasinska1982}, which we used in our fiducial model. At fixed $T_{\rm e}(\rm O^{++})$, our relation predicts a slightly lower $T_{\rm e}(\rm O^{+})$. However, this difference is smaller than the typical scatter in the relation. Our average relation is also closely aligned with the relation derived by \cite{cataldi2025}, derived using integrated galaxy spectra.

This large scatter is due to a combination of factors such as the different regions of ionised gas traced by these ions due to different ionisation potentials but also possible contamination such as the [\textsc{Oii}]$\lambda\lambda$7320,7330 measurements due to telluric emission lines. Additionally, temperature fluctuations and gradients within H\,\textsc{ii} regions and the presence of emission from multiple star-forming regions with different temperatures can further increase the observed dispersion \citep[see e.g.,][for detailed discussion on temperature measurements and also the resulting effect on abundance measurements]{peimbert1967,stasinska2005,bresolin2008}. These effects highlight the challenges in using a single temperature relation to infer $T_{\rm e}(\rm O^{++})$ from $T_{\rm e}(\rm O^{+})$ (and vice versa) and therefore, this emphasizes the importance of direct measurements of individual ionisation zones whenever possible.

\subsubsection{The $T_{\rm e}(\rm S^{++})$ versus $T_{\rm e}(\rm O^{+})$ relation}
The relation between $T_{\rm e}(\rm S^{++})$ and $T_{\rm e}(\rm O^{+})$ is less well studied in the literature. In Figure \ref{fig:t_siii_oii} we show the measured electron temperatures and the relation derived from our observations, which is given by
\begin{equation}
    T_{\rm e}({\rm O^{+}}) = (0.654\pm0.017) \times T_{\rm e}({\rm S^{++}}) + (0.326\pm0.018) \times 10^4~\rm K.
\end{equation}
The total scatter in $T_{\rm e}(\rm O^{+})$-residuals is $\sigma_{\rm tot}(\rm O^{+}) = 1400~\rm K$ and the intrinsic scatter is $\sigma_{\rm int}(\rm O^{+})=1000~\rm K$, where the residual is defined as $T_{\rm e}(\rm O^{+})_{\rm obs} - T_{\rm e}(\rm O^{+})_{\rm inferred}$. The total scatter in $T_{\rm e}(\rm S^{++})$ is $\sigma_{\rm tot}(\rm S^{++}) = 2100~\rm K$ and with intrinsic scatter $\sigma_{\rm int}(\rm S^{++})=1600~\rm K$. The relation we derive is slightly steeper than our fiducial model derived using the combination of the \cite{campbell1986} and \cite{garnett1992} and \cite{croxall2016} relations (see Fig. \ref{fig:t_siii_oii}). However, the differences are small in comparison to the typical scatter in the relation.

\begin{figure*}
	\begin{center}
	\includegraphics[width=\textwidth]{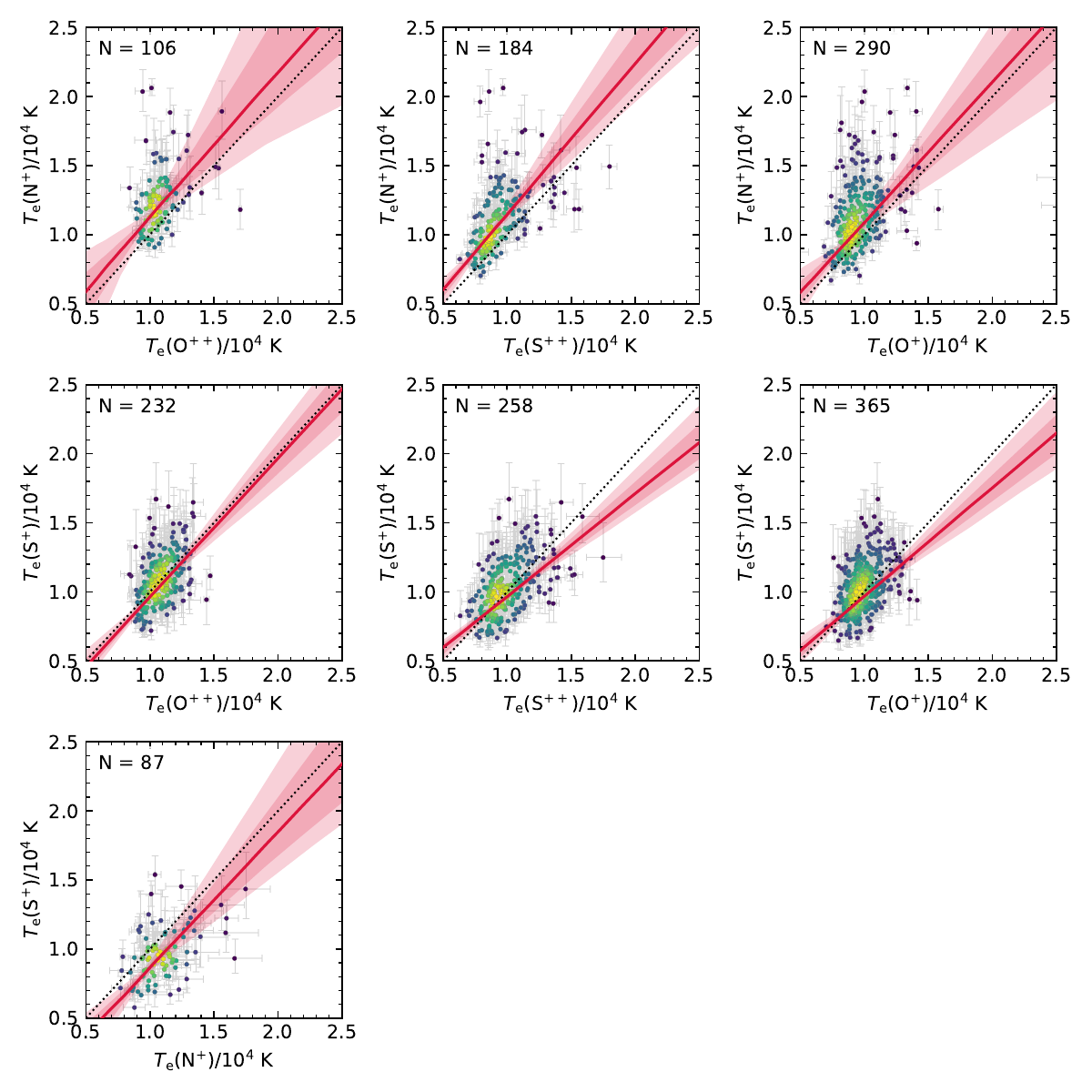}
	\caption{A compilation of the electron temperature relations between measurements from different ions. The red lines show the best fit relations derived from our observations. The shaded regions show the 1$\sigma$ uncertainties in the relations. All data points are coloured by the Gaussian kernel density of the plotted distribution.}
	\label{fig:temp_other}
	\end{center}
\end{figure*}

\subsubsection{Other temperature relations}
The remaining combinations of electron temperature measurements comprise significantly smaller samples. Therefore we do not discuss them in detail, however, the derived relations are provided below:
\begin{equation}
    T_{\rm e}({\rm N^{+}}) = (1.1\pm0.4) \times T_{\rm e}({\rm O^{++}}) - (0.0\pm0.4) \times 10^4~\rm K,
\end{equation} \vspace{-16pt}
\begin{equation}
    T_{\rm e}({\rm N^{+}}) = (1.11\pm0.16) \times T_{\rm e}({\rm S^{++}}) + (0.04\pm0.13) \times 10^4~\rm K,
\end{equation} \vspace{-16pt}
\begin{equation}
    T_{\rm e}({\rm N^{+}}) = (1.01\pm0.21) \times T_{\rm e}({\rm O^{+}}) + (0.08\pm0.18) \times 10^4~\rm K,
\end{equation} \vspace{-16pt}
\begin{equation}
    T_{\rm e}({\rm S^{+}}) = (1.01\pm0.11) \times T_{\rm e}({\rm O^{++}}) - (0.04\pm0.11) \times 10^4~\rm K,
\end{equation} \vspace{-16pt}
\begin{equation}
    T_{\rm e}({\rm S^{+}}) = (0.75\pm0.08) \times T_{\rm e}({\rm S^{++}}) + (0.21\pm0.08) \times 10^4~\rm K,
\end{equation} \vspace{-16pt}
\begin{equation}
    T_{\rm e}({\rm S^{+}}) = (0.79\pm0.10) \times T_{\rm e}({\rm O^{+}}) - (0.18\pm0.10) \times 10^4~\rm K,
\end{equation} \vspace{-16pt}
\begin{equation}
    T_{\rm e}({\rm S^{+}}) = (1.02\pm0.22) \times T_{\rm e}({\rm N^{+}}) - (0.15\pm0.22) \times 10^4~\rm K.
\end{equation}
The fitted relations are also shown in Figure \ref{fig:temp_other}. We note that the $T_{\rm e}({\rm N^{+}})$ temperatures of a significant fraction of galaxies is higher than expected assuming $\rm N^{+}$ and $\rm O^{+}$ trace the same gas conditions. This is visible in the number of high $T_{\rm e}({\rm N^{+}})$ outliers in the relevant panels of Figure \ref{fig:temp_other}. This is a similar finding to, e.g. \cite{arellano-cordova2024} where systematically high temperatures of $\rm N^{+}$ were also found. The $T_{\rm e}({\rm S^{+}})$ measurements are similar to $T_{\rm e}({\rm O^{+}})$, as is expected due to the similar ionisation potential of these ions. We note that particularly in these diagrams with smaller samples, the fitted relations do not always pass through the highest density region of the data. This is due to the fact that the ODR fitting includes the uncertainties in the measurements, where more uncertain measurements are given less weight in the line fitting.

\begin{figure*}
	\begin{center}
	\includegraphics[width=1.0\textwidth, trim=0 0.6cm 0 0]{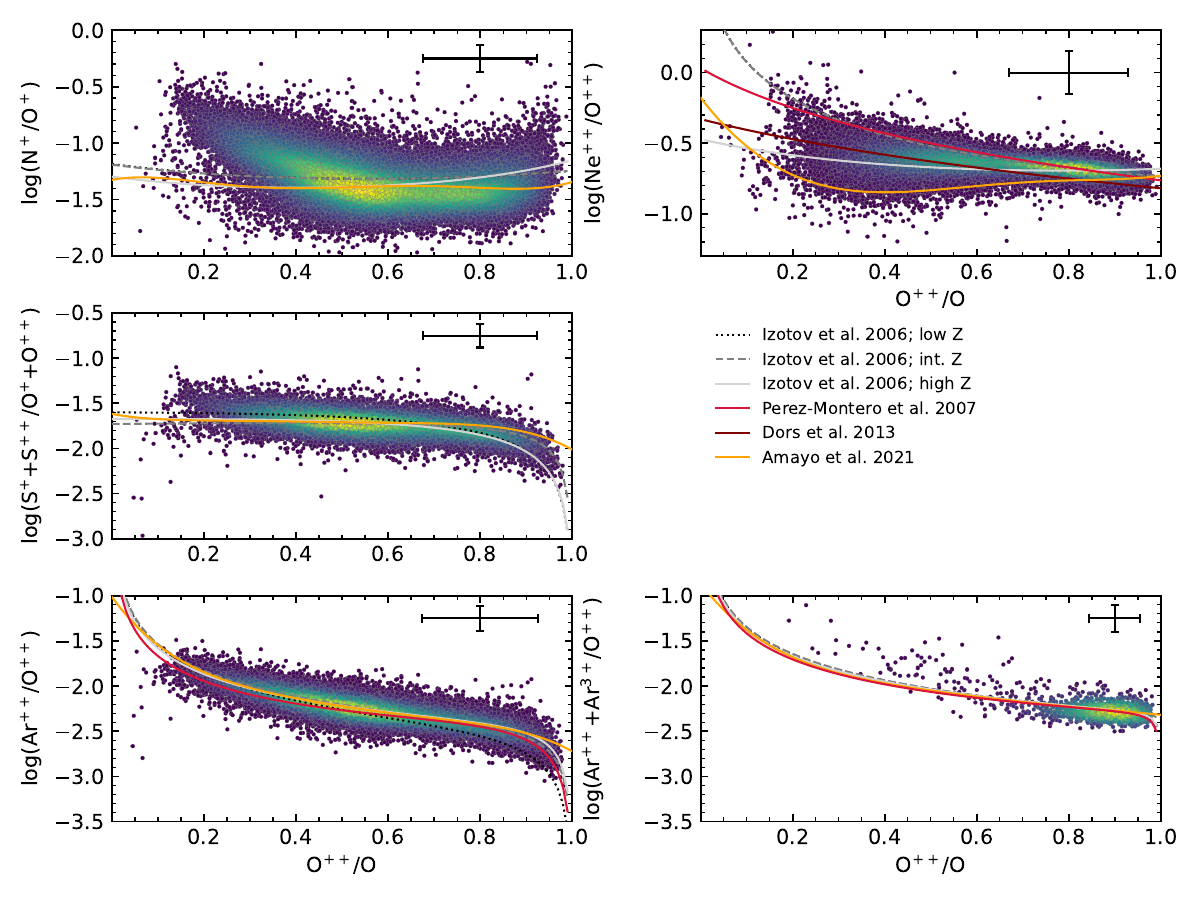}
	\caption{The ionic abundance ratios as a function of the degree of ionisation, defined as $\rm O^{++}/O$, shown together with the relations for ionisation correction factors by \protect\cite{izotov2006}, \protect\cite{perez-montero2007}, \protect\cite{dors2013} and \protect\cite{amayo2021} as indicated by the legend. Due to systematic evolution of the $\rm N/O$ ratio the data is not expected to coincide with the plotted ICFs in the top-left panel as those plotted lines are based on a constant N/O. The ICF functions are scaled to the median total abundance ratios of the plotted data using the \protect\cite{izotov2006} ICFs as described in Section \ref{sec:abundances}. All data points are coloured by the Gaussian kernel density of the plotted distribution. Typical uncertainties are shown by the black error bars in the top-right corner of each panel.}
	\label{fig:icfs}
	\end{center}
\end{figure*}

\subsection{Ionisation correction factors}
\label{sec:icfs}
We derive ionic abundances of the $\rm N^{+}$, $\rm O^{+}$, $\rm O^{++}$, $\rm Ne^{++}$, $\rm S^{+}$, $\rm S^{++}$, $\rm Ar^{++}$ and $\rm Ar^{3+}$ ions. To derive total elemental abundances we use ionisation correction factors to account for unobserved ionisation states. In Figure \ref{fig:icfs} we show the ionic abundance ratios as a function of the degree of ionisation, defined as $\rm O^{++}/(O^{+}+O^{++})$. We compare our measurements to the ICFs by \cite{izotov2006} which we use in our fiducial model. The ICFs by \cite{izotov2006} are based on photoionisation models and are widely used in the literature. We also compare to the ICFs derived by \cite{perez-montero2007}, \cite{dors2013} and \cite{amayo2021}. Due to systematic evolution of the $\rm N/O$ ratio the data is not expected to coincide with the plotted ICFs in the top-left panel as those plotted lines are based on a constant N/O. Typically, different ICFs are in close agreement with each other for N, S and Ar. Some of the largest deviations are observed for Ne at low degrees of ionisation where the different ICFs predict different contributions of unobserved Ne$^{+}$. Our observations generally cover the full range covered by the different ICFs, however, the high-metallicity ICF by \cite{izotov2006} provides a good match to the average of our measurements at low $\rm O^{++}/(O^{+}+O^{++})$, where most of the galaxies in this regime are in the high-metallicity selection of this ICF. We find that the total argon abundances are on average $0.13_{-0.09}^{+0.15}$ dex higher when the $\rm Ar^{3+}$ abundances are included. The $\rm Ar^{3+}$ abundances are only constrained for a small subset of 1$\,$260 out of 39$\,$692 galaxies with argon abundance measurements. The [Ar\textsc{iv}]$\lambda$4740 emission line is almost exclusively detected in galaxies with a high degree of ionisation (high $\rm O^{++}/O$) where a larger fraction of the Ar is in the Ar$^{3+}$ ionised state.

\subsection{Elemental abundances}
\label{sec:abundances}
In this section we present our total abundance measurements of N, O, Ne, S and Ar. Throughout this section we only present the abundances of galaxies where the $\rm log(X/H)^{84}- log(X/H)^{16}<0.6 ~dex$. Additionally, at $\rm 12+log(O/H) < 7.69~ dex$ we only report abundances where there is a constraint on the electron temperature from the [\textsc{Oiii}]$\lambda$4363 or [\textsc{Siii}]$\lambda$6312 auroral lines. See Table \ref{tab:sample_numbers} for a summary of the number of abundance measurements of each element.

\begin{figure*}
	\begin{center}
	\includegraphics[width=0.8\textwidth]{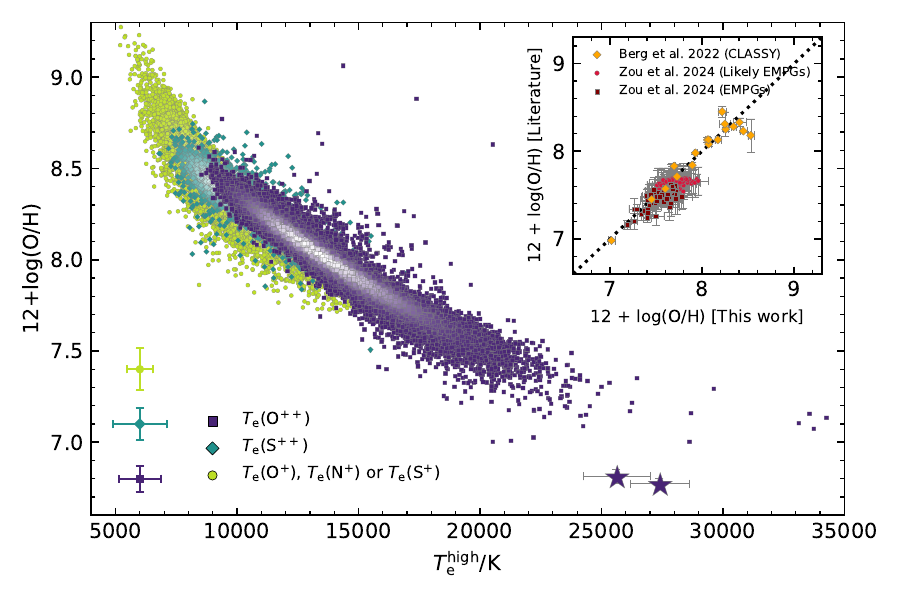}
	\caption{Oxygen abundances as a function of electron temperature of the high ionisation zone. We show $T_{\rm e}(\rm O^{++})$ (purple squares), for objects where this is measured, and show the converted values of $T_{\rm e}(\rm S^{++})$ (teal diamonds) or $T_{\rm e}(\rm O^{+})$, $T_{\rm e}(\rm N^{+})$ or $T_{\rm e}(\rm S^{+})$ (lime-green circles) using the fiducial temperature relations if the preceding are not available. The data points are shaded by the Gaussian kernel density of the plotted distributions. Typical uncertainties are shown by the coloured error bars in the bottom-left corner of each panel. We also highlight the two lowest metallicity galaxies in our sample, DESI J169.0571+14.1514 and DESI J211.9086+28.2461 (purple stars). For a detailed analysis of the lowest-metallicity galaxies in this sample we refer to Moustakas et al. (in prep.). The inset panel shows a comparison of our measurements to the direct abundance measurements of galaxies in the CLASSY survey \protect\citep{berg2022} and measurements of DESI extremely metal poor galaxies (EMPGs) and EMPG candidates in DESI early data by \protect\cite{zou2024}.}
	\label{fig:te_oh}
	\end{center}
\end{figure*}

\subsubsection{Oxygen abundances}
Oxygen is the most abundant metal in the ISM of galaxies and its abundance is typically used to quantify the gas-phase metallicity of galaxies. In Figure \ref{fig:te_oh} we show the oxygen abundance as a function of electron temperature of the high ionisation zone. There is a strong inverse relation between the oxygen abundance and electron temperature. This is mainly due to the increased efficiency of gas cooling in more metal rich gas. We show $T_{\rm e}(\rm O^{++})$, for objects where this is measured, and show the converted values of $T_{\rm e}(\rm S^{++})$, $T_{\rm e}(\rm O^{+})$, $T_{\rm e}(\rm N^{+})$ or $T_{\rm e}(\rm S^{+})$ using the fiducial temperature relations if the preceding are not available. 

Our measurements cover a wide range of oxygen abundances, from the lowest metallicity galaxies in the local Universe to high-metallicity galaxies with 99\% of our measurements between $\rm 7.42<12+log(O/H)<8.85$ dex. We generally find good agreement with literature measurements as shown in the inset panel in Figure \ref{fig:te_oh}. We compare to the direct abundance measurements of galaxies in the CLASSY survey \citep{berg2022} as well as measurements of DESI extremely metal poor galaxies (EMPGs) and EMPG candidates in DESI early data by \cite{zou2024}. These comparisons show that our measurements are consistent with previous literature. Due to the large number of galaxies observed by DESI, and the increased survey depth in comparison to e.g., the SDSS Main Galaxy Survey \citep{york2000}, we are able to detect a much larger number of galaxies with measurable auroral lines, and therefore direct-metallicity constraints, than has previously been possible. 

An example of this is the number of EMPGs detected in our dataset. Our sample contains 2$\,$855 galaxies with metallicities below 10\% solar, which is a significant increase in the known sample of EMPGs in the nearby Universe \citep[see e.g.,][for a detailed study of SDSS and DESI EMPGs]{sanchezamleida2016,zou2024}. This new sample will make more detailed studies of the chemical evolution of galaxies in pristine environments possible. This is important in particular to improve our understanding of the chemical evolution of galaxies in the early Universe, where the gas is known to be more metal-poor than in the local Universe on average \citep{sanders2021,jain2025,stanton2025b, zahid2011}. 

\begin{figure*}
	\begin{center}
	\includegraphics[width=\textwidth]{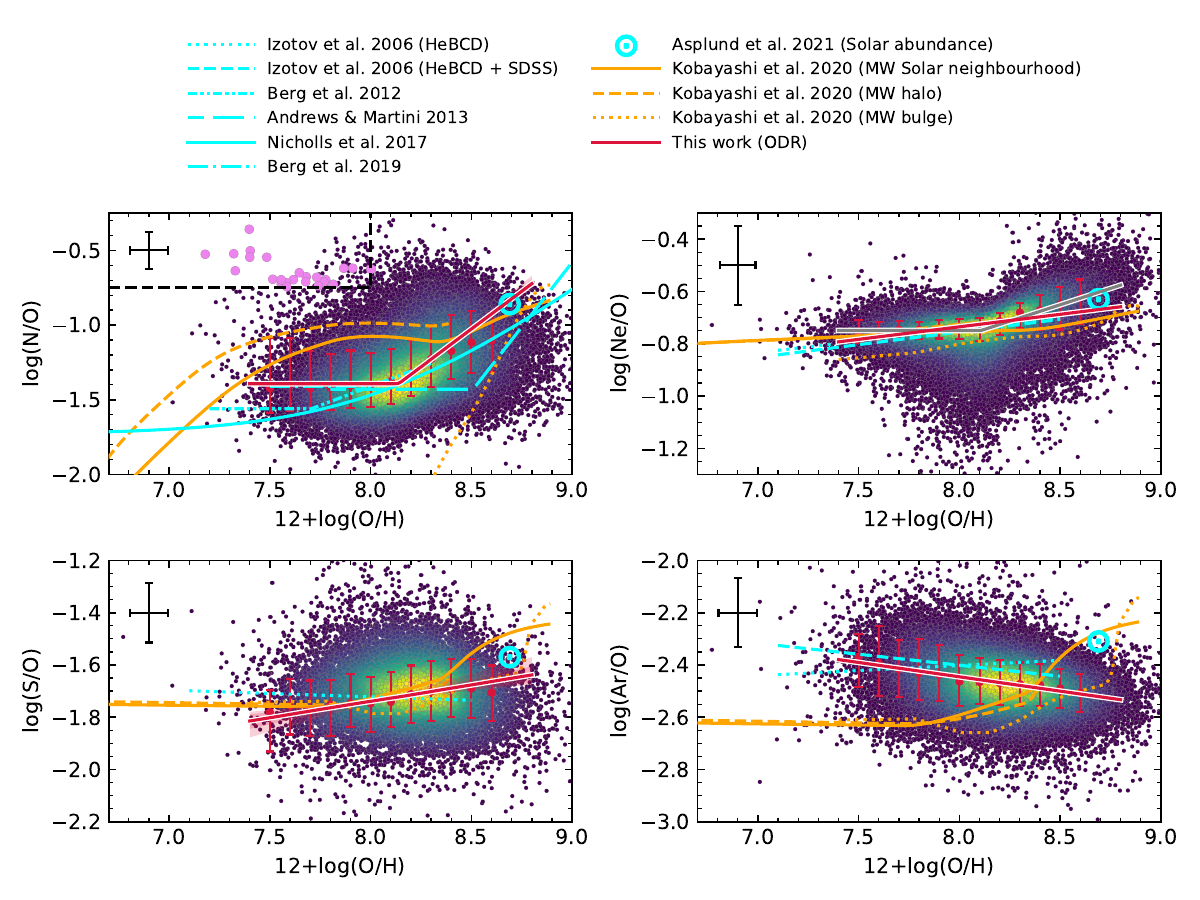}
	\caption{The abundance ratios, $\rm log(X/O)$, versus metallicity for nitrogen (top-left), neon (top-right), sulphur (bottom-left) and argon (bottom-right). The blue lines shows literature relations derived by \protect\cite{izotov2006}, \protect\cite{andrews2013}, \protect\cite{nicholls2017} and \protect\cite{berg2012,berg2019}. The orange lines show the predicted relations from chemical evolution models for the Solar neighbourhood, Milky Way halo and bulge stars by \protect\cite{kobayashi2020}. The cyan $\odot$-sign shows the solar abundance ratio \citep{asplund2021}. The pink datapoints in the top-left panel show a selection of galaxies with low metallicity and high $\rm N/O$ ratios as discussed in the text. All data points are coloured by the Gaussian kernel density of the plotted distribution. Typical uncertainties are shown by the black error bars in the top-left corner of each panel.}
	\label{fig:oh_vs_xo}
	\end{center}
\end{figure*}

We discover two galaxies with metallicities lower than any previous confirmed direct measurements to date. Our lowest metallicity galaxy, DESI J211.9086+28.2461, has an oxygen abundance of $\rm 12+\log(O/H) = 6.77_{-0.03}^{+0.03}~\rm dex$ or $\sim 1.2\% ~\rm Z_{\odot}$. We also highlight DESI J169.0571+14.1514, for which only the $\rm O^{++}$ abundance is measured due to a non-detection of the [\textsc{Oii}]$\lambda\lambda$3726,3729 doublet, suggesting a negligible fraction of the oxygen abundance in the $\rm O^{+}$ ionic state. Based on $\rm O^{++}$, we find an oxygen abundance of $\rm 12+\log(O/H) = 6.81_{-0.04}^{+0.04}~\rm dex$ or $\sim 1.3\% ~\rm Z_{\odot}$. The metallicities are lower than previous extremely metal-poor record holders with direct-metallicity measurements such as HSC J1631+4426 at $\rm 12+\log(O/H) = 6.90\pm 0.03~\rm dex$ \citep{kojima2020} in the local Universe and EXCELS-63107 at $\rm 12+\log(O/H) = 6.89_{-0.21}^{+0.26}~\rm dex$ \citep{cullen2025} in the early Universe \citep[there are some more metal-poor candidates at high-redshift, however, not confirmed by direct measurements due to the extreme observational challenge;][]{vanzella2025,cai2025, hsiao2025}. The detection of these galaxies highlights the power of the DESI survey to detect rare objects with extreme properties. For a more detailed analysis of the lowest-metallicity galaxies in this sample we refer to Moustakas et al. (in prep.).

\begin{figure*}
	\begin{center}
	\includegraphics[width=\textwidth]{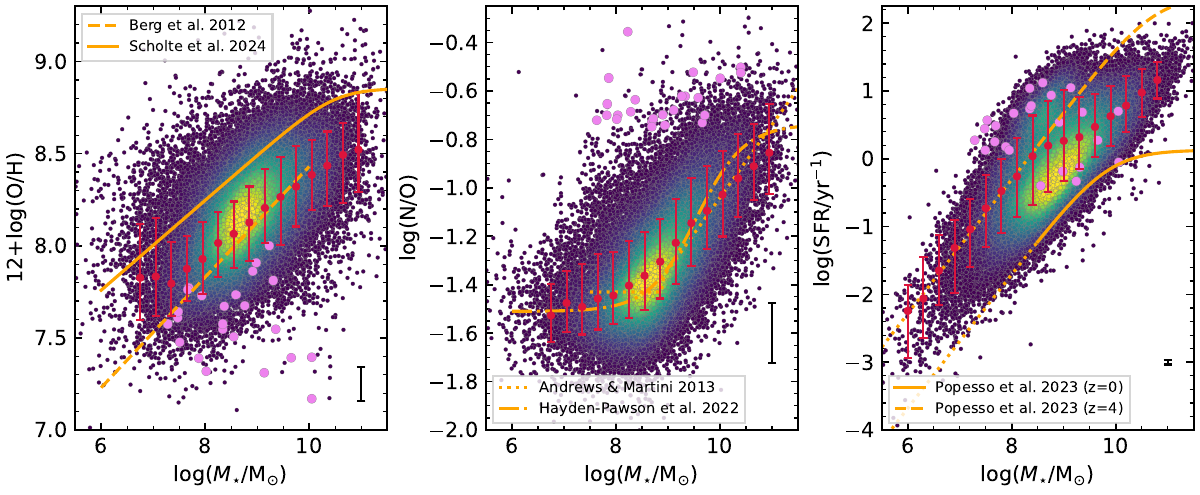}
	\caption{The mass-metallicity relation (left), the mass-$\rm N/O$ relation (middle) and the star formation main sequence (right). The data points are coloured by the Gaussian kernel density of the plotted distribution. Typical uncertainties are shown by the black error bars in the bottom-right corner of each panel. The red data with error bars show the median measurements and 16\numth to 84\numth percentiles at fixed stellar mass. The orange lines show measured relations from the literature as indicated in the legend \protect\citep{berg2012, andrews2013, hayden-pawson2022, popesso2023, scholte2024}. In the right panel the dotted lines indicate the region where the literature relation is extrapolated. The pink datapoints in the panels show a selection of galaxies with low metallicity and high $\rm N/O$ ratios as discussed in the text.}
	\label{fig:mstar_vs_no}
	\end{center}
\end{figure*}

Whilst the increased depth of DESI has resulted in a significant increase in the number of galaxies for which direct metallicities can be derived, it is still the case that the auroral lines are only detected in a small fraction of individual galaxy spectra. Due to this the sample of galaxies included here is not a representative sample of the galaxy population. Instead, the sample is biased towards strongly star forming, bright galaxies. These biases are reflected in the scaling relations of the galaxies in this sample, e.g., the mass-metallicity relation in the left panel of Fig. \ref{fig:mstar_vs_no} is biased towards systematically lower metallicities as compared to more representative samples \citep[see e.g.,][]{scholte2024}. In the right panel of Fig. \ref{fig:mstar_vs_no}, we also show that the galaxies in our sample have systematically higher star formation rates than typical $z=0$ galaxies; the average star formation rates are more similar to high redshift galaxies \citep[see e.g.,][]{popesso2023}. 

\subsubsection{Nitrogen abundances and ratios}
Contrary to oxygen, which is mostly produced in massive stars, the main production channel of nitrogen in galaxies is through intermediate mass stars \citep[e.g.,][]{kobayashi2020}. Due to these different production channels there is a large scatter in the observed nitrogen to oxygen abundance ratio, $\rm N/O$, in star-forming galaxies. In the top-left panel of Figure \ref{fig:oh_vs_xo} we show the $\rm N/O$ ratio as a function of oxygen abundance. We also note that the dynamic range is twice as large in the y-axis for the $\rm N/O$ panel than any of the other panels. Similar to previous literature, we fit a broken linear relation to the $\rm log(O/H)$ versus $\rm log(N/O)$ relation. We use the orthogonal distance regression as implemented in \textsc{Scipy} \citep[][]{boggs1981,scipy2020} to fit the relation. The uncertainty in the $\rm O/H$ and $\rm N/O$ measurements are included as the weights, $w$, in the ODR fitting, where $w_x=((\log(\rm O/H)^{84}-\log(\rm O/H)^{16})/2)^{-2}$ and $w_y=((\log(\rm N/O)^{84}-\log(\rm N/O)^{16})/2)^{-2}$. The fitted relation is given by
\begin{equation}
\label{eq:broken_linear}
	\log({\rm N/O}) = \begin{cases}
	c_{\rm plateau}, & \text{if } x < c_{\rm break} \\
	c_{\rm slope} ~ x + c_{\rm plateau}-c_{\rm slope}\times c_{\rm break}, & \text{if } x \geq c_{\rm break}
	\end{cases}
\end{equation}
where $x = 12+\log({\rm O/H})$, and the values of the fitting parameters are: $c_{\rm plateau} = -1.393\pm0.003$, $c_{\rm break} = 8.140\pm0.013$ and $c_{\rm slope} = 1.03\pm0.06$.
Alongside our measurements we also show the derived relations between $\rm O/H$ and $\rm N/O$ by \cite{berg2012}, \cite{andrews2013} and \cite{nicholls2017} as well as the $\rm N/O$ ratio of metal poor galaxies as determined by \cite{berg2019} and the solar abundance ratio \citep{asplund2021}. We also show the abundance ratios predicted from Milky Way chemical evolution models by \cite{kobayashi2020} for the Solar neighbourhood, the Milky Way bulge and halo stars. The average distribution of our measurements is in good agreement with the relations derived by \cite{berg2012} and \cite{nicholls2017} at $\rm 12+\log(O/H) > 8.0$, however, at lower metallicity we get better agreement with the relation of \cite{berg2019} who find $\rm \log(N/O) = -1.41\pm0.09$ dex at low-metallicity, where we find $\rm \log(N/O)=-1.393\pm0.003$ dex. This value is also in agreement with the $\rm \log(N/O)$ ratio of blue compact galaxies derived by \cite{izotov1999} at $-1.46\pm0.14$ dex. Generally, there is still some disagreement on the $\rm \log(N/O)$ plateau at low metallicities, with some studies finding a plateau at $\rm \log(N/O) \sim -1.4$ dex \citep[e.g.,][]{izotov1999,berg2019} and others finding a plateau at lower values of $\rm log(N/O) \sim -1.6$ dex \citep[e.g.,][]{berg2012,nicholls2017}. The higher plateau of our $\rm N/O$ ratio at lower metallicities could be due to sample incompleteness due to the faintness of the [\textsc{Nii}]$\lambda$6584 line at low metallicities, particularly for galaxies with low $\rm N/O$ ratios. However, as the higher plateau is also detected in the stacks of galaxy spectra from \cite{andrews2013} it may be a real feature of the galaxy population.

There are also differences in the literature about the slope of the O/H versus N/O relation in the high metallicity regime as well as the transition point where N/O starts to increase, e.g., at $8.142\pm0.014$ dex in this work compared to $\sim8.50$ dex in \cite{andrews2013}. These differences are most likely due to the dependencies of other parameters such as the stellar mass and SFR of the galaxies in the selected sample \citep[see also][]{boardman2024,boardman2024b,boardman2025}. In particular, the comparison to \cite{andrews2013} highlights the effect of sample selection. The detection requirement of auroral emission lines in individual galaxies means that our sample contains more strongly star forming galaxies (as shown in the right panel of Fig. \ref{fig:mstar_vs_no}), whereas \cite{andrews2013} are sensitive to a more representative sample of galaxies as they use stacked measurements to detect auroral emission lines. Therefore, at fixed metallicity their sample will include more faint, low SFR and low mass galaxies which have systematically lower N/O (see middle panel of Fig. \ref{fig:mstar_vs_no} in this work, or Fig. 14 in \citealp{andrews2013}).

The chemical evolution models reveal that the $\rm N/O$ ratio is strongly dependent on the modelled region in the MW chemical evolution models of \cite{kobayashi2020}. This is a reflection of the different nucleosynthesis processes and distribution timescales into the ISM of these elements. The observed scatter in the $\rm N/O$ ratio at fixed metallicity is likely due to variations in the star formation history of galaxies. More detailed modelling of the chemical evolution is able to reproduce the observed relation based on these processes, see e.g., \cite{vincenzo2018a} and \cite{vincenzo2018b}.

\begin{figure*}
	\begin{center}
	\includegraphics[width=\textwidth, trim=0 0.9cm 0 0.2cm, clip]{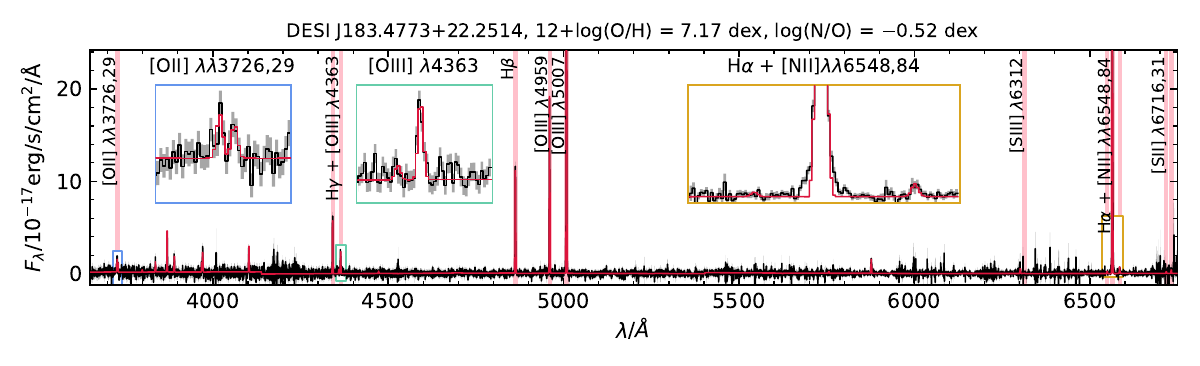}
    \includegraphics[width=\textwidth, trim=0 0.9cm 0 0.2cm, clip]{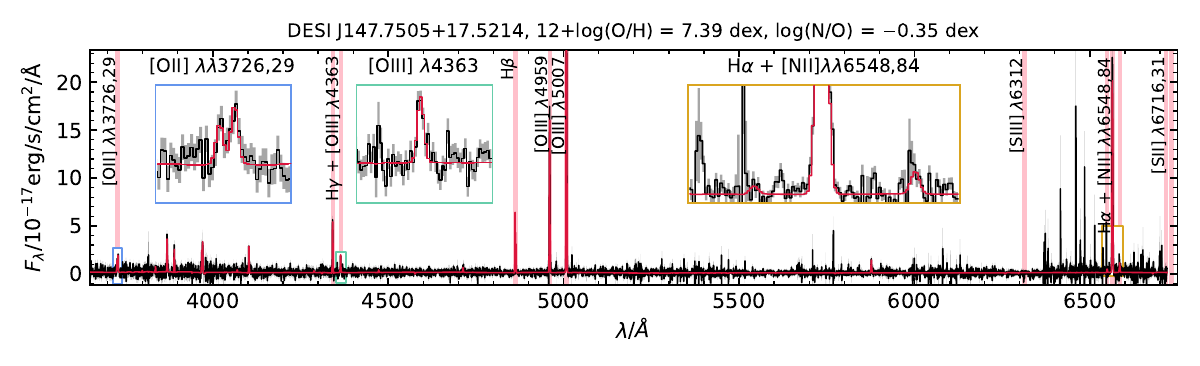}
	\includegraphics[width=\textwidth, trim=0 0.9cm 0 0.2cm, clip]{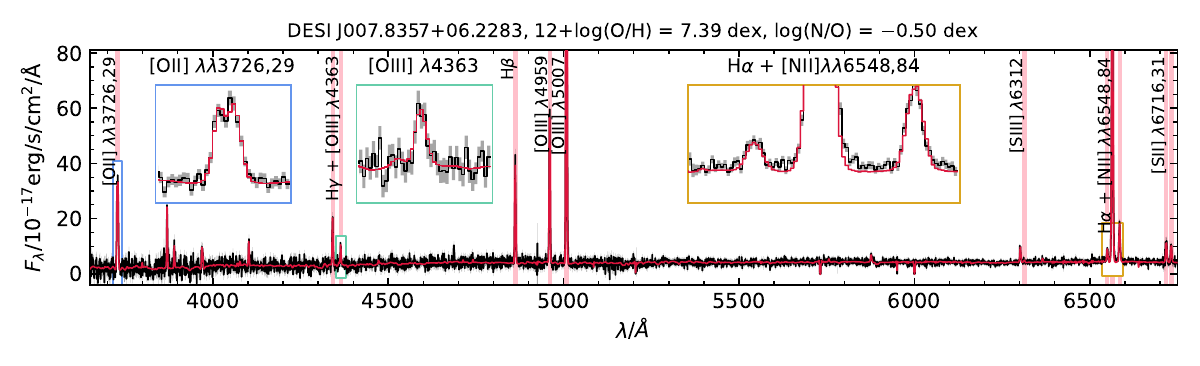}	\includegraphics[width=\textwidth, trim=0 0 0 0.2cm, clip]{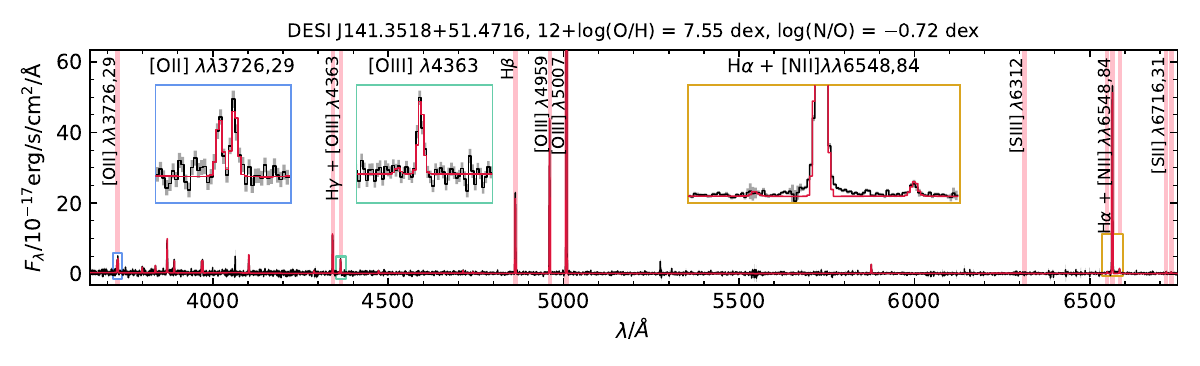}
	\caption{Example spectra of galaxies with high N/O abundance ratios. Each panel shows inset zoom-ins of the [\textsc{Oii}]$\lambda\lambda$3726,3729, [\textsc{Oiii}]$\lambda$4363 and H$\alpha$ + [\textsc{Nii}]$\lambda\lambda$658
    48,6584 emission lines. \textit{Panel 1:} This shows the most metal poor high N/O galaxy in our sample. \textit{Panel 2:} The galaxy with the highest N/O ratio. The nitrogen line of this galaxy falls in a region with large sky subtraction spikes which may contribute to an artificially high [NII]$\lambda$6584 flux. \textit{Panel 3:} One of the high N/O galaxies falling into the composite region of the [NII]-BPT diagram. \textit{Panel 4:} A galaxy close to the high-N/O selection boundary.}
	\label{fig:high_no_spectra}
	\end{center}
\end{figure*}

We show the mass-metallicity relation (MZR), mass-$\rm N/O$ relation and star formation main sequence in Figure \ref{fig:mstar_vs_no}. Similar to other studies we find a strong relation between the stellar mass and the oxygen abundance, with more massive galaxies having higher oxygen abundances \citep[see e.g.,][]{tremonti2004, berg2012,andrews2013}. We find that at fixed stellar mass our sample has systematically lower metallicities than the general population of galaxies with strong emission line detections \citep[as shown by the MZR of DESI DR1 galaxies derived using strong line calibrations that are calibrated to the direct metallicity scale;][]{scholte2024}, which is expected from the selection criteria which preferentially select more strongly star-forming galaxies where auroral emission lines are detected. The $\rm N/O$ ratio is also strongly correlated with the stellar mass. Our sample shows a similar trend to the relations derived by \cite{andrews2013} and \cite{hayden-pawson2022}. 

There is significant interest in nitrogen-rich galaxies due to the detection of high $\rm \log(N/O)$ galaxies in the early Universe such as GN-z11 with $\rm log(N/O)>-0.25$ dex \citep[][]{cameron2023}, CEERS-1018 with $\rm log(N/O) = -0.18\pm0.11$ dex \citep{marques-chaves2024} and RXCJ2248-ID with $\rm log(N/O)=-0.39_{-0.08}^{+0.10}$ dex \citep{topping2024}. Our sample contains a significant number of galaxies with high $\rm \log(N/O)$ ratios at low $\rm \log(O/H)$. We selected a sample of 103 outliers at $\rm 12+\log(O/H) < 8.0$ and $\rm \log(N/O) > -0.75$. We further subselect to only include galaxies with a constraint on the electron temperature from the [\textsc{Oiii}]$\lambda$4363 or [\textsc{Siii}]$\lambda$6312 auroral lines. This final sample of 24 nitrogen-rich galaxies is shown by the pink datapoints and outlined parameter space in the top-left panel of Figure \ref{fig:oh_vs_xo}. We exclude galaxies without [\textsc{Oiii}]$\lambda$4363 or [\textsc{Siii}]$\lambda$6312 temperature constraints to minimise the effect of scatter in the electron temperature relations which can lead to underestimated oxygen abundances and overestimated N/O ratios. The high metallicity selection boundary is chosen to exclude the metallicity regime where high N/O is common. We choose the N/O threshold to select galaxies with N/O similar to extreme nitrogen emitters observed in the early Universe. 

We find that the high N/O galaxies selected are typically more massive than the average population of galaxies at similar metallicity (albeit with a large scatter; see Figure \ref{fig:mstar_vs_no}). This is consistent with findings by \cite{arellano-cordova2025} who find that high $\rm \log(N/O)$ galaxies are more massive than typical low-metallicity galaxies. They find that this is the case for both local and high-redshift galaxies \citep[see also][]{bhattacharya2025desi}. This finding is generally consistent with the interpretation of gas inflows diluting the metallicity, resulting in a reduction of O/H at fixed N/O \citep[e.g.,][]{andrews2013, curti2020, scholte2023, scholte2024, scholte2025}. Many but not all of the selected high N/O galaxies are significantly more star forming than other galaxies in our sample at the same mass as would be expected if gas dilution is the mechanism behind the high-N/O outliers. It is possible that the high N/O galaxies are tracing an early period of star formation shortly after the infall of pristine gas where the metallicity has been reduced but the high N/O abundance ratio, produced by AGB stars after the previous burst of star formation, has been retained \citep[see e.g.,][]{mcclymont2025}. The question still remains whether the high N/O abundance pattern is indeed an observed phase after significant gas accretion or whether there is another mechanism producing this peculiar abundance pattern. An alternative may be the rapid production and release of large amounts of nitrogen by Wolf-Rayet stars \citep{berg2011, rivera-thorsen2024, berg2025} or even very massive stars \citep{vink2023} following a burst of star formation. 

It should also be considered that it is possible that the derived peculiar abundance patterns are the result of physical processes that are not considered in the analysis such as AGN contamination or very high electron density ionised gas. Indeed, 5/24 of our high N/O galaxies reside in the composite region of the [NII]-BPT diagram where they are classified as star forming according to the \cite{kewley2001} criterion but not \cite{kauffmann2003}. Additionally, when examining outlier populations in a large statistical dataset it is important to consider possible contamination in the spectra (e.g., due to sky subtraction errors) or the need for more sophisticated multiple component line measurements. As shown in Figure \ref{fig:high_no_spectra}, several of our high-N/O galaxies show faint broad components that are not fully captured by our FastSpecFit model. These could be fitted in more detail using bespoke line fitting, which is beyond the scope of this work. We can also see that the [\textsc{Nii}]$\lambda$6584 line of our highest-N/O candidate is located in a region with large sky subtraction spikes which may boost the flux of this emission line. Nonetheless, upon visual inspection the spectra and line fitting generally are robust in this outlier population. Therefore showing that, whilst rare, high N/O abundances at low metallicity do indeed exist even at low redshift.

\subsubsection{Neon abundances and ratios}
As neon and oxygen are both alpha-elements, their abundances are expected to be strongly correlated. In the top-right panel of Figure \ref{fig:oh_vs_xo} we show the $\rm Ne/O$ ratio as a function of oxygen abundance. We find that the median $\rm log(Ne/O)$ ratio is $-0.74_{-0.08}^{+0.11}$ dex. We fit a linear function to the relation using orthogonal distance regression:
\begin{equation}
	\log({\rm Ne/O}) = (0.0957\pm0.0020) x - (1.501\pm0.016),
\end{equation}
where $x=12+\log(\rm O/H)$. This relation is consistent with the relation derived by \cite{izotov2006} and with the expected relation derived from chemical evolution models \citep{kobayashi2020}. The average $\rm O/Ne$ ratio is also consistent with the solar value \citep{asplund2021} at solar metallicities.

At metallicities higher than $\rm 12+\log(O/H) \sim 8.1$, we find an increasing trend in the $\rm Ne/O$ ratio, compared to a mostly flat trend at lower metallicities. Therefore, we also fit the $\rm Ne/O$ ratio using a broken linear relation (see Eq. \ref{eq:broken_linear}) providing the following fitted parameters: $c_{\rm plateau} = -0.7496\pm0.0006$, $c_{\rm break} = 8.105\pm0.004$ and $c_{\rm slope} = 0.250\pm0.004$. The best fit relation is plotted in grey in Figure \ref{fig:oh_vs_xo}. This result is similar to the findings by e.g., \cite{amayo2021}, \cite{miranda-perez2023}, \cite{arellano-cordova2024} and \cite{esteban2025}, who each also find a more sharply increasing trend. 

The nature for the increasing trend in $\rm Ne/O$ at high metallicity is still unclear. As shown in Section \ref{sec:icfs}, there is a large spread in the ionisation correction factors of neon which complicates the interpretation of these results. The differences between these ICFs is large enough to suggest that the sharp increase in the $\rm Ne/O$ ratio could be due to inaccurate ICFs. However, this trend could also reflect real changes in the $\rm Ne/O$ abundance ratio of the ionized gas at high metallicity. This could be due to the increased depletion of oxygen onto dust grains at higher metallicities, which would lead to a higher gas phase $\rm Ne/O$ ratio. Dust-to-metal ratios in the ISM of galaxies are expected to increase with metallicity \citep[e.g.,][]{galliano2018}, which would lead to a higher depletion of oxygen onto dust grains at higher metallicities. Particularly, dust evolution models and observations suggest that the dust-to-metal ratio may increase sharply at $\rm 12+log(O/H)\sim 8.2$ dex \citep{devis2017, devis2019}. This would lead to a higher $\rm Ne/O$ ratio in the gas phase at high metallicities, as neon is not depleted onto dust grains but oxygen is depleted. Compared to a dust free scenario the expected increase in the $\rm \log(Ne/O)$ ratio is $\sim0.2$ dex, assuming typical ISM depletion fractions of oxygen \citep{jenkins2009}. This is consistent with the observed increase in the $\rm \log(Ne/O)$ ratio at high metallicities.

\subsubsection{Sulphur and argon abundances and ratios}
Similar to neon, the elements sulphur and argon are alpha-elements, therefore, their abundances are expected to be strongly correlated with the oxygen abundance. However, there is a more complex set of production pathways of these elements, where a significant fraction of the S and Ar is produced in SNe Ia \citep[e.g.,][]{kobayashi2020}. This leads to a more complex relation between the sulphur, argon abundances and the oxygen abundances as has been confirmed observationally in both planetary nebulae \citep{arnaboldi2022} and star-forming galaxies \citep[e.g.,][]{stanton2025,bhattacharya2025,bhattacharya2025jwst}. 

In the bottom-left panel of Figure \ref{fig:oh_vs_xo} we show the $\rm S/O$ ratio as a function of oxygen abundance. We find a median $\rm log(S/O)$ ratio of $-1.70_{-0.11}^{+0.12}$ dex. We fit a linear function to the relation using orthogonal distance regression:
\begin{equation}
	\log({\rm S/O}) = (0.14\pm0.04) x - (2.82\pm0.03),
\end{equation}
where $x=12+\log(\rm O/H)$. This is broadly consistent with the relation derived by \cite{izotov2006}, although with a different slope, and in agreement with the expected relation derived from chemical evolution models \citep{kobayashi2020}. Our average $\rm S/O$ ratio at solar metallicities is lower than the solar value \citep{asplund2021}, however, this is within the scatter of our measurements.

In the bottom-right panel of Figure \ref{fig:oh_vs_xo} we show the $\rm Ar/O$ ratio as a function of oxygen abundance. We find the $\rm log(Ar/O)$ ratio has a median value of $-2.48_{-0.11}^{+0.11}$ dex. We fit a linear function to the relation using orthogonal distance regression:
\begin{equation}
	\log({\rm Ar/O}) = -(0.110\pm0.005) x - (1.57\pm0.04),
\end{equation}
where $x=12+\log(\rm O/H)$. Our average $\rm Ar/O$ ratio is similar to the values found by \cite{izotov2006}, and \cite{arellano-cordova2024}. However, particularly at low metallicity, our average $\rm Ar/O$ ratio is significantly higher than the expected relation derived from chemical evolution models \citep{kobayashi2020}. Similar to our results for sulphur, our average $\rm Ar/O$ ratio at solar metallicities is lower than the solar value \citep{asplund2021}, however, within the scatter of our measurements. 

\begin{figure}
	\begin{center}
	\includegraphics[width=\columnwidth]{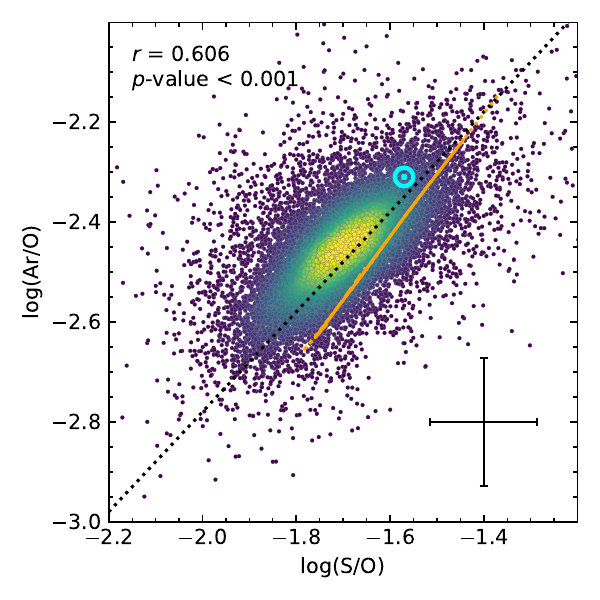}
	\caption{The $\rm S/O$ ratio as a function of the $\rm Ar/O$ ratio. The dashed black line shows the expected relation with unity slope corrected for the relative difference in abundance between sulphur and argon as measured through the median values of our sample: $\rm log(Ar/O) = log(S/O) - 0.78 ~dex$. The strong correlation between the two ratios is expected due to the similar production channels of sulphur and argon, this is also shown by the tracks of chemical evolution models in orange, as indicated by the legend on Figure \ref{fig:oh_vs_xo}. The pearson correlation coefficient and associated $p$-value are shown in the top-left corner of the panel. The orange lines show the predicted relations from chemical evolution models for the Solar neighbourhood, Milky Way halo and bulge stars by \protect\cite{kobayashi2020}. All data points are coloured by the Gaussian kernel density of the plotted distribution. Typical uncertainties are shown by the black error bars in the bottom-right corner of each panel.}
	\label{fig:so_vs_aro}
	\end{center}
\end{figure}

Finally, as both sulphur and argon have a production channel through SNe Ia, the $\rm S/O$ and $\rm Ar/O$ abundance ratios are expected to be correlated. In Figure \ref{fig:so_vs_aro} we show that this is indeed the case. Where the dashed black line is the expected relation with unity slope corrected for the relative difference in abundance between sulphur and argon as measured through the median values of our sample: $\rm log(Ar/O) = log(S/O) -0.78 ~dex$. We also show the Pearson correlation coefficient  between the two ratios, which is $r=0.606$ with a $p$-value less than $10^{-3}$. This indicates a significant correlation.

\section{Summary and conclusions} \label{sec:summary}
In this work we presented a comprehensive analysis of the electron temperatures and abundances of N, O, Ne, S and Ar in star-forming galaxies using the DESI DR2 dataset. The main findings of our analysis are summarised below:

\begin{itemize}
	\item Our dataset contains electron temperature and abundance measurements for 49$\,$959 galaxies. The exact breakdown of the number of galaxies with electron temperature and abundance measurements is shown in Table \ref{tab:sample_numbers}.
	\item We find that the average relation between $T_{\rm e}(\rm O^{++})$ and $T_{\rm e}(\rm S^{++})$ is close to the one-to-one relation. This suggests that the electron temperatures derived from these ions likely trace similar physical conditions in the ionized gas (Figure \ref{fig:t_oiii_siii}).
	\item We find a large scatter in the relation between $T_{\rm e}(\rm O^{++})$ and $T_{\rm e}(\rm O^{+})$, which is consistent with previous literature (Figure \ref{fig:t_oiii_oii}). 
	\item We provide constraints on a number of other temperature relations as discussed in Section \ref{sec:tt_relations}. 
	\item We find a large sample of 2$\,$855 extremely metal poor galaxies (EMPGs) with metallicities less than 10\% solar. These EMPGs are ideal candidates for studying the chemical evolution of galaxies in pristine environments such as present in the early Universe.
	\item We discover two of the most metal-poor galaxies in the local Universe, DESI J211.9086+28.2461 and DESI J169.0571+14.1514, with oxygen abundances of $\rm 12+\log(O/H) = 6.77_{-0.03}^{+0.03}~\rm dex$ and $\rm 12+\log(O/H) = 6.81_{-0.04}^{+0.04}~\rm dex$, respectively. For a detailed analysis of the lowest-metallicity galaxies in this sample we refer to Moustakas et al. (in prep.).
	\item We find that the average $\rm N/O$ ratio of the galaxies in our sample is consistent with previous literature results \citep[e.g.,][]{izotov1999,berg2012, nicholls2017, berg2019}. 
    \item We also find a population of galaxies with high $\rm N/O$ ratios at low $\rm O/H$. There are several possible scenarios for the physical nature of this population. This will be investigated in more detail in a further study (Scholte et al. in prep.).
	\item We find that the average $\rm Ne/O$ ratio is consistent with the relation derived by \cite{izotov2006} and with the expected relation derived from chemical evolution models \citep{kobayashi2020}. However, we also find a more sharply increasing trend in the $\rm Ne/O$ ratio at high metallicities. This elevated trend may be caused by systematic effects in the measurements or could reflect real changes in the physical conditions of the ionized gas at high metallicities. A likely explanation for this trend is the increased depletion of oxygen onto dust grains at higher metallicities, which would lead to a higher gas phase $\rm \log(Ne/O)$ ratio.
	\item We find that the average $\rm S/O$ ratio is consistent with the relation from chemical evolution models \citep{kobayashi2020}, however, $\rm Ar/O$ shows some discrepancies, particularly at low metallicity. We find a strong correlation between the $\rm S/O$ and $\rm Ar/O$ ratios, which is expected due to the similar production channels of these elements.
\end{itemize}

This new dataset provides a wealth of information on the physical conditions and chemical abundances in star-forming galaxies. The number of galaxies with auroral line detections allows us to perform statistical studies of the chemical evolution of galaxies, including outlier populations, which has previously been very challenging. We will study the abundance relations of galaxies including outlier populations in more detail in a follow-up paper (Scholte et al. in prep.). 
There are several other key areas where this dataset will allow significant progress. Electron temperature relations may be improved by including additional properties \citep[as shown by e.g.,][]{langeroodi2024}. This will benefit from large number of galaxies with multiple abundance measurements as included in our DESI DR2 dataset. The derivation of robust strong line metallicity calibrations that account for secondary dependencies also requires reliable measurements of metallicities of large samples of galaxies \citep[see also e.g.,][]{nakajima2022}. As shown by \cite{scholte2025}, once secondary dependencies are accounted for, these strong line metallicity calibrations will be applicable at any redshift. 

The $\sim2$ orders of magnitude step change in the number of individual galaxies for which we can measure auroral emission lines is due to the increased survey depth and number of galaxies observed by DESI. As DESI continues its 8-year survey we may double the size of this sample once again. Therefore, DESI will remain the largest statistical sample of galaxy abundances for the foreseeable future and be a core resource for the study of the chemical evolution of galaxies. It is already providing a valuable local comparison for chemical evolution studies at high-redshift using JWST \citep[see e.g.,][]{scholte2025, laseter2025} and will provide a core reference for chemical evolution studies in upcoming surveys such as MOONS (Multi-Object Optical and Near-infrared Spectrograph), PFS (Prime Focus Spectrograph), WEAVE (WHT Enhanced Area Velocity Explorer) and 4MOST (4-metre Multi-Object Spectrograph Telescope).

\section*{Acknowledgements}
The authors thank the anonymous journal reviewer for their thoughtful comments which have improved the quality of this work. We also thank the DESI internal review panel, particularly Alejandro Aviles, Paul Martini and Kelly Douglass, for their valuable comments and suggestions that have improved the quality of this work. We would like to thank Ragadeepika Pucha for helpful suggestions on emission line fitting.

D. Scholte, F. Cullen, K. Z. Arellano-C\'ordova and T. M. Stanton and acknowledge support from a UKRI Frontier Research Guarantee Grant (PI Cullen; grant reference EP/X021025/1).

This material is based upon work supported by the U.S. Department of Energy (DOE), Office of Science, Office of High-Energy Physics, under Contract No. DE–AC02–05CH11231, and by the National Energy Research Scientific Computing Center, a DOE Office of Science User Facility under the same contract. Additional support for DESI was provided by the U.S. National Science Foundation (NSF), Division of Astronomical Sciences under Contract No. AST-0950945 to the NSF’s National Optical-Infrared Astronomy Research Laboratory; the Science and Technology Facilities Council of the United Kingdom; the Gordon and Betty Moore Foundation; the Heising-Simons Foundation; the French Alternative Energies and Atomic Energy Commission (CEA); the National Council of Humanities, Science and Technology of Mexico (CONAHCYT); the Ministry of Science, Innovation and Universities of Spain (MICIU/AEI/10.13039/501100011033), and by the DESI Member Institutions: \url{https://www.desi.lbl.gov/collaborating-institutions}. Any opinions, findings, and conclusions or recommendations expressed in this material are those of the author(s) and do not necessarily reflect the views of the U. S. National Science Foundation, the U. S. Department of Energy, or any of the listed funding agencies.

The authors are honored to be permitted to conduct scientific research on I'oligam Du'ag (Kitt Peak), a mountain with particular significance to the Tohono O’odham Nation.

The DESI Legacy Imaging Surveys consist of three individual and complementary projects: the Dark Energy Camera Legacy Survey (DECaLS), the Beijing-Arizona Sky Survey (BASS), and the Mayall z-band Legacy Survey (MzLS). DECaLS, BASS and MzLS together include data obtained, respectively, at the Blanco telescope, Cerro Tololo Inter-American Observatory, NSF’s NOIRLab; the Bok telescope, Steward Observatory, University of Arizona; and the Mayall telescope, Kitt Peak National Observatory, NOIRLab. NOIRLab is operated by the Association of Universities for Research in Astronomy (AURA) under a cooperative agreement with the National Science Foundation. Pipeline processing and analyses of the data were supported by NOIRLab and the Lawrence Berkeley National Laboratory (LBNL). Legacy Surveys also uses data products from the Near-Earth Object Wide-field Infrared Survey Explorer (NEOWISE), a project of the Jet Propulsion Laboratory/California Institute of Technology, funded by the National Aeronautics and Space Administration. Legacy Surveys was supported by: the Director, Office of Science, Office of High Energy Physics of the U.S. Department of Energy; the National Energy Research Scientific Computing Center, a DOE Office of Science User Facility; the U.S. National Science Foundation, Division of Astronomical Sciences; the National Astronomical Observatories of China, the Chinese Academy of Sciences and the Chinese National Natural Science Foundation. LBNL is managed by the Regents of the University of California under contract to the U.S. Department of Energy. The complete acknowledgments can be found at https://www.legacysurvey.org/acknowledgment/.

The Siena Galaxy Atlas was made possible by funding support from the U.S. Department of Energy, Office of Science, Office of High Energy Physics under Award Number DE-SC0020086 and from the National Science Foundation under grant AST-1616414.

\section*{Data Availability}

The data from DESI Data Release 2 is not yet publicly available. However, the data from the DESI Survey Validation period and Data Release 1 is publicly available at \url{https://data.desi.lbl.gov/doc/}. This includes spectra and derived data such as emission line flux measurements from \textsc{FastSpecFit} \citep{moustakas2023}. The Legacy Survey imaging is available at \url{https://www.legacysurvey.org/}. After the public release of DESI Data Release 2, the derived electron temperature and abundance measurements will be made available. A link to the data will also be shared at \url{https://dirkscholte.github.io/data}. A summary of our data model is already available in appendix \ref{sec:datamodel}. Data shown in figures will be available on Zenodo upon acceptance (\url{https://zenodo.org/records/21414309}). Please note that whilst a significant amount of time was spent vetting the measurements, the size of the dataset required emission line fitting using an automated pipeline. Not every spectrum or fitted emission line has been individually inspected. This should be noted when considering the properties of individual galaxies.



\bibliographystyle{mnras}
\bibliography{main} 



\appendix

\section{Data model}\label{sec:datamodel}
The measurements derived in this work comprise a catalogue that will be made available after the public release of DESI DR2. The data model for this catalogue is provided in Table \ref{tab:datamodel}.
\renewcommand{\arraystretch}{1.25}
\begin{table*}
\centering
\caption{The data model of the measurement catalogue presented in this work.}
\begin{tabular}{lll}
\hline
\textbf{Quantity} & \textbf{Units} & \textbf{Description} \\
\hline \hline
DESINAME & --- & The official DESI name of the galaxy \\
TARGETID & --- & Unique target ID \\
SURVEY & --- & Survey name \\
PROGRAM & --- & Programme name \\
HEALPIX & --- & Healpixel number \\
RA & deg & Right ascension \\
DEC & deg & Declination \\
Z & --- & Redshift \\
LOGMSTAR & $\rm M_{\odot}$ & The stellar mass computed using \textsc{FastSpecFit} assuming $h=0.6766$. \\
LOGSFR\_P[$n$] & $\rm M_{\odot}~yr^{-1}$ & The [$n=16,50,84$]\numth percentile value of star formation rate based on dust corrected H$\alpha$ or H$\beta$ line emission. \\
AV\_P[$n$] & mag & The [$n=16,50,84$]\numth percentile value of the dust attenuation measurement. \\
NE\_OII\_P[$n$] & cm$^{-3}$ & The [$n=16,50,84$]\numth percentile value of the electron density. \\
TE\_OIII\_P[$n$] & K & The [$n=16,50,84$]\numth percentile value of the $\rm O^{++}$ electron temperature. \\
TE\_SIII\_P[$n$] & K & The [$n=16,50,84$]\numth percentile value of the $\rm S^{++}$ electron temperature. \\
TE\_OII\_P[$n$] & K & The [$n=16,50,84$]\numth percentile value of the $\rm O^{+}$ electron temperature. \\
TE\_NII\_P[$n$] & K & The [$n=16,50,84$]\numth percentile value of the $\rm N^{+}$ electron temperature. \\
TE\_SII\_P[$n$] & K & The [$n=16,50,84$]\numth percentile value of the $\rm S^{+}$ electron temperature. \\
N+\_P[$n$] & dex & The [$n=16,50,84$]\numth percentile value of the $\rm log(N^{+}/H^{+})$ abundance. \\
NH\_P[$n$] & dex & The [$n=16,50,84$]\numth percentile value of the $\rm log(N/H)$ abundance. \\
O+\_P[$n$] & dex & The [$n=16,50,84$]\numth percentile value of the $\rm log(O^{+}/H^{+})$ abundance. \\
O++\_P[$n$] & dex & The [$n=16,50,84$]\numth percentile value of the $\rm log(O^{++}/H^{+})$ abundance. \\
OH\_P[$n$] & dex & The [$n=16,50,84$]\numth percentile value of the $\rm log(O/H)$ abundance. \\
NE++\_P[$n$] & dex & The [$n=16,50,84$]\numth percentile value of the $\rm log(Ne^{++}/H^{+})$ abundance. \\
NEH\_P[$n$] & dex & The [$n=16,50,84$]\numth percentile value of the $\rm log(Ne/H)$ abundance. \\
S+\_P[$n$] & dex & The [$n=16,50,84$]\numth percentile value of the $\rm log(S^{+}/H^{+})$ abundance. \\
S++\_P[$n$] & dex & The [$n=16,50,84$]\numth percentile value of the $\rm log(S^{++}/H^{+})$ abundance. \\
SH\_P[$n$] & dex & The [$n=16,50,84$]\numth percentile value of the $\rm log(S/H)$ abundance. \\ 
AR++\_P[$n$] & dex & The [$n=16,50,84$]\numth percentile value of the $\rm log(Ar^{++}/H^{+})$ abundance. \\
AR+++\_P[$n$] & dex & The [$n=16,50,84$]\numth percentile value of the $\rm log(Ar^{3+}/H^{+})$ abundance. \\
ARH\_P[$n$] & dex & The [$n=16,50,84$]\numth percentile value of the $\rm log(Ar/H)$ abundance. \\ 
\hline
\end{tabular}
\label{tab:datamodel}
\end{table*}
\renewcommand{\arraystretch}{1.}

\section{Author affiliations}

$^{1}$Institute for Astronomy, University of Edinburgh, Royal Observatory, Blackford Hill, Edinburgh EH9 3HJ, UK\\
$^{2}$Department of Physics and Astronomy, Siena College, 515 Loudon Road,
Loudonville, NY 12110, USA\\
$^{3}$National Astronomical Observatories, Chinese Academy of Sciences, A20
Datun Road, Chaoyang District, Beijing 100012, China\\
$^{4}$Department of Physics \& Astronomy, University College London, Gower Street, London, WC1E 6BT, UK\\
$^{5}$Max-Planck Institute for Radio Astronomy, Auf dem Hugel ¨ 69, D-53121
Bonn, Germany\\
$^{6}$Department of Physics and Astronomy and PITT PACC, University of
Pittsburgh, Pittsburgh, PA 15260, USA\\
$^{7}$Lawrence Berkeley National Laboratory, 1 Cyclotron Road, Berkeley, CA 94720, USA\\
$^{8}$Department of Physics, Boston University, 590 Commonwealth Avenue, Boston, MA 02215 USA\\
$^{9}$Dipartimento di Fisica ``Aldo Pontremoli'', Universit\`a degli Studi di Milano, Via Celoria 16, I-20133 Milano, Italy\\
$^{10}$INAF-Osservatorio Astronomico di Brera, Via Brera 28, 20122 Milano, Italy\\
$^{11}$Institut d'Estudis Espacials de Catalunya (IEEC), c/ Esteve Terradas 1, Edifici RDIT, Campus PMT-UPC, 08860 Castelldefels, Spain\\
$^{12}$Institute of Space Sciences, ICE-CSIC, Campus UAB, Carrer de Can Magrans s/n, 08913 Bellaterra, Barcelona, Spain\\
$^{13}$Kapteyn Astronomical Institute, University of Groningen, Landleven 12 (Kapteynborg, 5419), 9747 AD Groningen, The Netherlands\\
$^{14}$Instituto de F\'{\i}sica, Universidad Nacional Aut\'{o}noma de M\'{e}xico,  Circuito de la Investigaci\'{o}n Cient\'{\i}fica, Ciudad Universitaria, Cd. de M\'{e}xico  C.~P.~04510,  M\'{e}xico\\
$^{15}$Department of Astronomy \& Astrophysics, University of Toronto, Toronto, ON M5S 3H4, Canada\\
$^{16}$Department of Physics \& Astronomy, University of Rochester, 206 Bausch and Lomb Hall, P.O. Box 270171, Rochester, NY 14627-0171, USA\\
$^{17}$University of California, Berkeley, 110 Sproul Hall \#5800 Berkeley, CA 94720, USA\\
$^{18}$Departamento de F\'isica, Universidad de los Andes, Cra. 1 No. 18A-10, Edificio Ip, CP 111711, Bogot\'a, Colombia\\
$^{19}$Observatorio Astron\'omico, Universidad de los Andes, Cra. 1 No. 18A-10, Edificio H, CP 111711 Bogot\'a, Colombia\\
$^{20}$Institute of Cosmology and Gravitation, University of Portsmouth, Dennis Sciama Building, Portsmouth, PO1 3FX, UK\\
$^{21}$University of Virginia, Department of Astronomy, Charlottesville, VA 22904, USA\\
$^{22}$Fermi National Accelerator Laboratory, PO Box 500, Batavia, IL 60510, USA\\
$^{23}$NSF NOIRLab, 950 N. Cherry Ave., Tucson, AZ 85719, USA\\
$^{24}$Sorbonne Universit\'{e}, CNRS/IN2P3, Laboratoire de Physique Nucl\'{e}aire et de Hautes Energies (LPNHE), FR-75005 Paris, France\\
$^{25}$Center for Cosmology and AstroParticle Physics, The Ohio State University, 191 West Woodruff Avenue, Columbus, OH 43210, USA\\
$^{26}$Department of Astronomy, The Ohio State University, 4055 McPherson Laboratory, 140 W 18th Avenue, Columbus, OH 43210, USA\\
$^{27}$The Ohio State University, Columbus, 43210 OH, USA\\
$^{28}$Instituci\'{o} Catalana de Recerca i Estudis Avan\c{c}ats, Passeig de Llu\'{\i}s Companys, 23, 08010 Barcelona, Spain\\
$^{29}$Institut de F\'{i}sica d’Altes Energies (IFAE), The Barcelona Institute of Science and Technology, Edifici Cn, Campus UAB, 08193, Bellaterra (Barcelona), Spain\\
$^{30}$Department of Physics and Astronomy, University of Waterloo, 200 University Ave W, Waterloo, ON N2L 3G1, Canada\\
$^{31}$Perimeter Institute for Theoretical Physics, 31 Caroline St. North, Waterloo, ON N2L 2Y5, Canada\\
$^{32}$Waterloo Centre for Astrophysics, University of Waterloo, 200 University Ave W, Waterloo, ON N2L 3G1, Canada\\
$^{33}$Space Sciences Laboratory, University of California, Berkeley, 7 Gauss Way, Berkeley, CA  94720, USA\\
$^{34}$Instituto de Astrof\'{i}sica de Andaluc\'{i}a (CSIC), Glorieta de la Astronom\'{i}a, s/n, E-18008 Granada, Spain\\
$^{35}$Departament de F\'isica, EEBE, Universitat Polit\`ecnica de Catalunya, c/Eduard Maristany 10, 08930 Barcelona, Spain\\
$^{36}$Department of Physics and Astronomy, Sejong University, 209 Neungdong-ro, Gwangjin-gu, Seoul 05006, Republic of Korea\\
$^{37}$CIEMAT, Avenida Complutense 40, E-28040 Madrid, Spain\\
$^{38}$Department of Astronomy, School of Physics and Astronomy, Shanghai Jiao Tong University, Shanghai 200240, China\\
$^{39}$University of Michigan, 500 S. State Street, Ann Arbor, MI 48109, USA\\


\bsp	
\label{lastpage}
\end{document}